\pdfoutput=1

\documentclass[11pt]{article}

\usepackage[final]{acl}

\usepackage{times}
\usepackage{latexsym}

\usepackage[T1]{fontenc}

\usepackage[utf8]{inputenc}

\usepackage{microtype}

\usepackage{inconsolata}

\usepackage{graphicx}

\usepackage{amssymb}
\usepackage{amsmath}
\usepackage{booktabs}
\usepackage{multirow}

\usepackage{multirow}
\usepackage{enumitem}
\usepackage{tabularx}

\newcommand{\ourdataset}[1]{}
\renewcommand{\ourdataset}[1]{\textsc{MisBot}}

\title{How Do Social Bots Participate in Misinformation Spread?\\
A Comprehensive Dataset and Analysis}

\author{Herun Wan\textsuperscript{1, 2} \ \ \ \ \ \ \
Minnan Luo\textsuperscript{1, 2, *} \ \ \ \ \ \ \
Zihan Ma\textsuperscript{1, 3} \\ \bf
Guang Dai\textsuperscript{4} \ \ \ \ \ \ \
Xiang Zhao\textsuperscript{5} \\
\textsuperscript{1}School of Computer Science and Technology, Xi’an Jiaotong University, China\\ 
\textsuperscript{2}Ministry of Education Key Laboratory of Intelligent Networks and Network Security, China\\
\textsuperscript {3}Shaanxi Province Key Laboratory of Big Data Knowledge Engineering, China\\
\textsuperscript{4}SGIT AI Lab, State Grid Corporation of China\\
\textsuperscript{5}National University of Defense Technology \\
\href{mailto:wanherun@stu.xjtu.edu.cn}{\texttt{wanherun@stu.xjtu.edu.cn}}\ \ \ \ \ \href{mailto:minnluo@xjtu.edu.cn}{\texttt{minnluo@xjtu.edu.cn}}\\
\href{https://whr000001.github.io/MisBot/}{https://whr000001.github.io/MisBot/}
}

\begin{document}
\maketitle
\def\thefootnote{*}\footnotetext{Corresponding author}\def\thefootnote{\arabic{footnote}}

\begin{abstract}
Social media platforms provide an ideal environment to spread misinformation, where social bots can accelerate the spread. 
This paper explores the interplay between social bots and misinformation on the Sina Weibo platform. 
We construct a large-scale dataset that includes annotations for both misinformation and social bots. 
From the misinformation perspective, the dataset is multimodal, containing 11,393 pieces of misinformation and 16,416 pieces of verified information. 
From the social bot perspective, this dataset contains 65,749 social bots and 345,886 genuine accounts, annotated using a weakly supervised annotator. 
Extensive experiments demonstrate the comprehensiveness of the dataset, the clear distinction between misinformation and real information, and the high quality of social bot annotations. 
Further analysis illustrates that: (\romannumeral 1) social bots are deeply involved in information spread; (\romannumeral 2) misinformation with the same topics has similar content, providing the basis of echo chambers, and social bots would amplify this phenomenon; and (\romannumeral 3) social bots generate similar content aiming to manipulate public opinions.
\end{abstract}

\section{Introduction}
\begin{figure}[t]
    \centering
    \includegraphics[width=\linewidth]{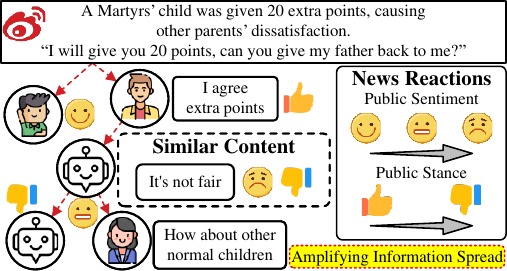}
    \caption{An example of social bots participating in information spread. Social bots would publish similar content to manipulate public sentiment and stance, leading to a shift in public opinion.}
    \label{fig: teaser}
\end{figure}

Social media platforms, like $\mathbb{X}$ (Twitter) and Weibo, have become major information sources, and information spreads faster than traditional media. Due to the nature of such platforms, there have been attempts to disseminate misinformation, which could polarize society \citep{azzimonti2023social} and impact the economy \citep{zhou2024characterizing}. Meanwhile, besides attracting genuine users, the social platform also becomes an ideal breeding ground for malicious social bots \citep{cresci2020decade} due to the straightforward operation. Social bots are proven behind many online perils, including election interference \citep{ng2022cross} and hate speech propaganda \citep{stella2019influence}. Social bots are natural message amplifiers \citep{caldarelli2020role}, increasing the risk of spreading misinformation \citep{huang2022social}. Namely, misinformation and social bots are two major factors harming online security. They might work together to amplify negative impact, where Figure \ref{fig: teaser} presents an example.

Researchers make efforts to fight the never-ending plague of misinformation and malicious social bots. 
They mainly propose automatic detectors to identify misinformation \citep{shu2019defend} and social bots \citep{yang2022botometer}. 
Meanwhile, researchers also explore how different types of content \citep{nan2021mdfend} or propagation patterns \citep{vosoughi2018spread} influence misinformation spread. 
From the social bot perspective, bot communities \citep{tan2023botpercent} and bots' repost behaviors \citep{elmas2022characterizing} have been investigated. 
While many works have provided valuable insights into investigating misinformation and social bots, relatively little attention \citep{wang2018era, himelein2021bots} has been paid to the interplay between them. 

This paper aims to bridge the gap of existing works, exploring the interplay between misinformation and social bots. We propose \ourdataset{}\footnote{The main language of \ourdataset{} is Chinese.}, a dataset which simultaneously contains information and annotations of misinformation and social bots (\S \ref{sec: dataset}). Specifically, we first define the structure of \ourdataset{}. We then collect misinformation from Weibo’s official management center\footnote{\url{https://service.account.weibo.com/}, being available for users who have logged in.}. After that, we collect real information from two credible sources to ensure the authenticity. We finally propose a weakly supervised annotator to label the users involved in the dissemination of information. From the misinformation perspective, \ourdataset{} contains multiple modalities, including post content, comments, repost messages, images, and videos. \ourdataset{} includes 11,393 misinformation instances and 16,416 real information instances. From the social bot perspective, \ourdataset{} includes 952,955 users participating in the information spread, covering 65,749 annotated social bots and 345,886 genuine accounts. Extensive experiments (\S \ref{sec: basic analysis}) prove that (\romannumeral 1) \ourdataset{} is the most comprehensive and the only one with misinformation and social bot annotations, (\romannumeral 2) misinformation and real information are distinguishable, where a simple detector achieves 95.2\% accuracy, and (\romannumeral 3) \ourdataset{} has high social bot annotation quality, where human evaluations prove it. Further analysis illustrates (\S \ref{sec: analysis}) that (\romannumeral 1) social bots are deeply involved in information spread, where 29.3\% users who repost misinformation are social bots; (\romannumeral 2) misinformation with the same topics has similar content, providing the basis of echo chambers, and social bots amplify this phenomenon; and (\romannumeral 3) social bots generate similar content aiming to manipulate public opinions, including sentiments and stances.

\section{\ourdataset{} Dataset}
\label{sec: dataset}
The collection process of \ourdataset{} consists of four components: (\romannumeral 1) \textbf{Data Structure} defines the dataset structure; (\romannumeral 2) \textbf{Misinformation Collection} collects multiple modalities in misinformation; (\romannumeral 3) \textbf{Real Information Collection} collects real information from two sources; and (\romannumeral 4) \textbf{Weakly Supervised User Annotation} trains a weakly supervised annotator to automatically annotate accounts.

\subsection{Data Structure}
Users publish posts to spread information on the Weibo platform, thus, we annotate user posts as misinformation or real information. From this perspective, each instance is represented as $A=\{s, \mathcal{G}_\textit{repost}, \mathcal{G}_\textit{comment}, I, V, y\}$. It contains textual content $s$, repost graph $\mathcal{G}_\textit{repost}$, comment graphs $\mathcal{G}_\textit{comment}$, images $I$, videos $V$, and corresponding label $y$. From the account perspective, each instance is represented as $U=\{F, T, y\}$. It contains the attribute set $F$, published posts $T$, and the corresponding label $y$. Meanwhile, we believe a user participates in the spread of a post if this user reposts, comments, or likes this post. Some cases in \ourdataset{} are provided in Appendix \ref{subapp: dataset example}.

\subsection{Misinformation Collection}
We collect posts flagged as misinformation from Weibo's official management center, where we provide the platform overview in Appendix \ref{subapp: management center} for readers who cannot log in. This platform presents posts containing misinformation judged by platform moderators or police. Besides, it provides a brief \textbf{judgment} to explain why the post is flagged as misinformation, which provides a basis for identifying topics of misinformation. An example is provided in Appendix \ref{subapp: misinformation example}. We have collected all the misinformation since this platform was established. Specifically, the misinformation collected was published between April 2018 and April 2024. We spent about 10 months collecting 11,393 pieces of misinformation.

\subsection{Real Information Collection}
Existing misinformation datasets generally suffer from potential data bias \citep{chen2023causal}, especially entity biases \citep{zhu2022generalizing}. It means that the entity distributions in misinformation and real information differ, influencing models' generalization ability to unseen data. Thus, we design an entity debiasing method to mitigate entity biases. We first employ a keyphrase extractor \citep{liang2021unsupervised} to obtain key entities from each misinformation. After filtering uncommon entities, we get 1,961 entities, where we present the filter rules in Appendix \ref{subapp: entity filter}. We finally query the key entities using the Weibo search engine in trusted sources to get real information. An overview of the search engine is provided in Appendix \ref{subapp: query method} for readers who cannot log in. To ensure the authenticity and diversity of real information, we collect real information from two sources: 
\begin{itemize}[topsep=4pt, leftmargin=*]
    \item \textbf{Verified news media} is an official news account certified by the Weibo platform, which contains a red ``verified'' symbol and a verified reason, where we provide the statistics of the verified news accounts in Appendix \ref{subapp: verified accounts}.
    \item \textbf{Trends on the platform} contains posts sparking a lot of discussion in a short period.
\end{itemize}

Due to the moderation of Weibo, we assume these two sources are truthful, where we discuss it further in Appendix \ref{subapp: source credibility} and quantitatively prove it in \S \ref{sec: basic analysis}. We obtained 8,317 and 8,099 pieces of real information, respectively.

\subsection{Weakly Supervised User Annotation}
Manual annotation or crowd-sourcing is labor-intensive and not feasible with large-scale datasets. Meanwhile, to ensure the scalability of \ourdataset{}, we propose a weakly supervised learning strategy, enabling automatic annotation. The construction of the weakly supervised annotator contains (\romannumeral 1) preprocessing, (\romannumeral 2) training, and (\romannumeral 3) inference phases.

\paragraph{Preprocessing Phase} This phase aims to collect the training dataset for the weakly supervised annotator. We first collected 100,000 random accounts. Due to the randomness, these accounts could represent the entire Weibo environment, ensuring the diversity of accounts. We employ crowd-sourcing to annotate them, where the human annotators are familiar with social media. Following existing works \citep{feng2021twibot, feng2022twibot}, we summarize a brief criteria for identifying a social bot on Weibo and write a guideline document for human annotators, where we provide the document in Appendix \ref{subapp: annotation guideline}. Notably, social bot annotation is subjective, where the average Fleiss’ Kappa is 0.4281 as shown in \S \ref{sec: basic analysis}. Thus, we do not directly define what a social bot is, but only provide a brief guideline document and cases. Inspired by existing work \citep{feng2021twibot}, we determine 20 standard accounts that are easy to identify. Each annotator should also annotate 20 standard accounts, and annotators who achieve more than 80\% accuracy on standard accounts are reliable. We ensure that each account is annotated by three reliable human annotators. We totally recruited 315 annotators and spent 60,000 yuan and 60 days, where we provide details in Appendix \ref{subapp: annotation cost}. We employ major voting to obtain the final annotations in this phase. 

Based on human annotators' feedback, we filter in active accounts in \ourdataset{}, where we provide the filter rules in Appendix \ref{subapp: active accounts}. We focus on active accounts for three reasons:
\begin{itemize}[topsep=4pt, leftmargin=*]
    \item We aim to explore the involvement of social bots in misinformation and real information spread, where inactive users hardly participate in information spread.
    \item Annotators mainly rely on posts in users' timelines to make judgments, whereas inactive accounts cannot provide enough information to obtain reliable annotations.
    \item Mainstream social bot detectors analyze accounts' posts to identify bots, and we follow this to construct an annotation model. We employ active users to ensure credibility.
\end{itemize}

We obtained 48,536 active accounts from the 100,000 accounts, of which 18,132 are social bots and 30,404 are genuine accounts.

\paragraph{Training Phase}
Different machine bot detectors have their strengths and weaknesses in the face of multiple social bots \citep{sayyadiharikandeh2020detection}. Thus, we propose to employ multiple detectors as experts and employ an ensemble strategy to obtain the final annotations. In this phase, we leverage the following detectors:
\begin{itemize}[topsep=4pt, leftmargin=*]
    \item \textbf{Feature-based detectors} leverage feature engineering on user attributes and adopt classic machine learning algorithms to identify social bots. We employ various attributes: (\romannumeral 1) \textbf{numerical}: \emph{follower count}, \emph{following count}, and \emph{status count}; and (\romannumeral 2) \textbf{categorical}: \emph{verified}, \emph{svip}, \emph{account type}, and \emph{svip level}. We employ MLP layers, random forests, and Adaboost as detectors.
    \item \textbf{Content-based detectors} encode user-generated textual content, where we employ \emph{name}, \emph{description}, and \emph{posts}. We employ encoder-based language models, including BERT \citep{devlin2019bert} and DeBERTa \citep{DBLP:conf/iclr/HeLGC21} to obtain textual representations and employ an MLP layer to identify social bots.
    \item \textbf{Ensemble detectors} concatenate the attribute and textual representations and employ an MLP layer to identify social bots.  
\end{itemize}
The descriptions and settings of experts are provided in Appendix \ref{subapp: expert settings}. We create an 8:1:1 split for the users from the preprocessing phase as train, validation, and test sets to train each expert.

\begin{table*}[t]
    \centering
    \setlength{\tabcolsep}{1.5mm} 
    \resizebox{\linewidth}{!}{
    \begin{tabular}{lcccccccccccc}
        \toprule[1.5pt]
        \multirow{2}{*}{\textbf{Dataset}}&\multicolumn{6}{c}{\textbf{Modalities}}&\multicolumn{6}{c}{\textbf{Statistics}}\\\cmidrule(lr){2-7}\cmidrule(lr){8-13}
        &Content&Comment&Repost&Image&Video&User&Post&Image&Video&User&Bots&Human\\
        \midrule[1pt]
        \multicolumn{13}{l}{\emph{Datasets for misinformation detection.}}\\
        \citep{shu2020fakenewsnet}$^\star$&\checkmark&\checkmark&\checkmark&\checkmark&&\checkmark&23,196&19,200&0&\textbf{2,063,442}&0&0\\
        \citep{nan2021mdfend}$^\star$&\checkmark&&&&&&9,128&0&0&0&0&0\\
        \citep{li2022cnn}&&&&&\checkmark&&700&0&700&0&0&0\\
        \citep{qi2023fakesv}$^\star$&&\checkmark&&&\checkmark&\checkmark&3,654&0&3,654&3,654&0&0\\
        \citep{hu2023mr2}&\checkmark&\checkmark&&\checkmark&\checkmark&&14,700&14,700&0&0&0&0\\
        \citep{li2024mcfend}$^\star$&\checkmark&\checkmark&&\checkmark&&\checkmark&23,789&10,178&0&803,779&0&0\\
        \midrule[1pt]
        \multicolumn{13}{l}{\emph{Datasets for social bot detection.}}\\
        \citep{feng2021twibot}&&&&&&\checkmark&0&0&0&229,580&6,589&5,237\\
        \citep{feng2022twibot}&&\checkmark&\checkmark&&&\checkmark&0&0&0&1,000,000&\textbf{139,943}&\textbf{860,057}\\
        \citep{shi2023mgtab}$^\star$&&\checkmark&\checkmark&&&\checkmark&0&0&0&410,199&2,748&7,451\\
        \midrule[1pt]
        \multicolumn{13}{l}{\emph{Datasets for the interplay between misinformation and social bots.}}\\
        \textbf{\ourdataset{}}&\checkmark&\checkmark&\checkmark&\checkmark&\checkmark&\checkmark&\textbf{27,809}&\textbf{61,714}&\textbf{7,328}&952,955&65,749&345,886\\
        \bottomrule[1.5pt]
    \end{tabular}
    }
    \caption{Summary of our dataset and recent datasets for misinformation and social bots. We first check each dataset's modality and then report the related statistics. The $\star$ denotes that the publisher does not provide the original data in the corresponding paper. Our dataset is the largest and the only one with misinformation and social bot annotations, containing 27,809 instances.}
    \label{tab: summary}
\end{table*}

\paragraph{Inference Phase} This phase annotates accounts based on the predictions from multiple experts. To ensure annotation quality, we filter in the experts achieving 80\% accuracy, which is the same standard as human annotators, on the validation set. After that, a conventional method to integrate multiple predictions is to employ majority voting or train an MLP classifier on the validation set \citep{bach2017snorkel, feng2022twibot}. Since the likelihood from classifiers may not accurately reflect true probabilities \citep{guo2017calibration}, also known as \emph{miscalibrated}, we calibrate the likelihoods before the ensemble. We employ temperature scaling \citep{guo2017calibration} and select the best temperature on the validation set, where we provide the temperature settings in Appendix \ref{subapp: temperature settings}. We finally average the calibrated likelihoods to obtain the final annotations. Among the 952,955 accounts that participate in information spread in \ourdataset{}, 411,635 are active, of which 65,749 are social bots and 345,886 are genuine accounts.

\begin{figure}[t]
    \centering
    \includegraphics[width=\linewidth]{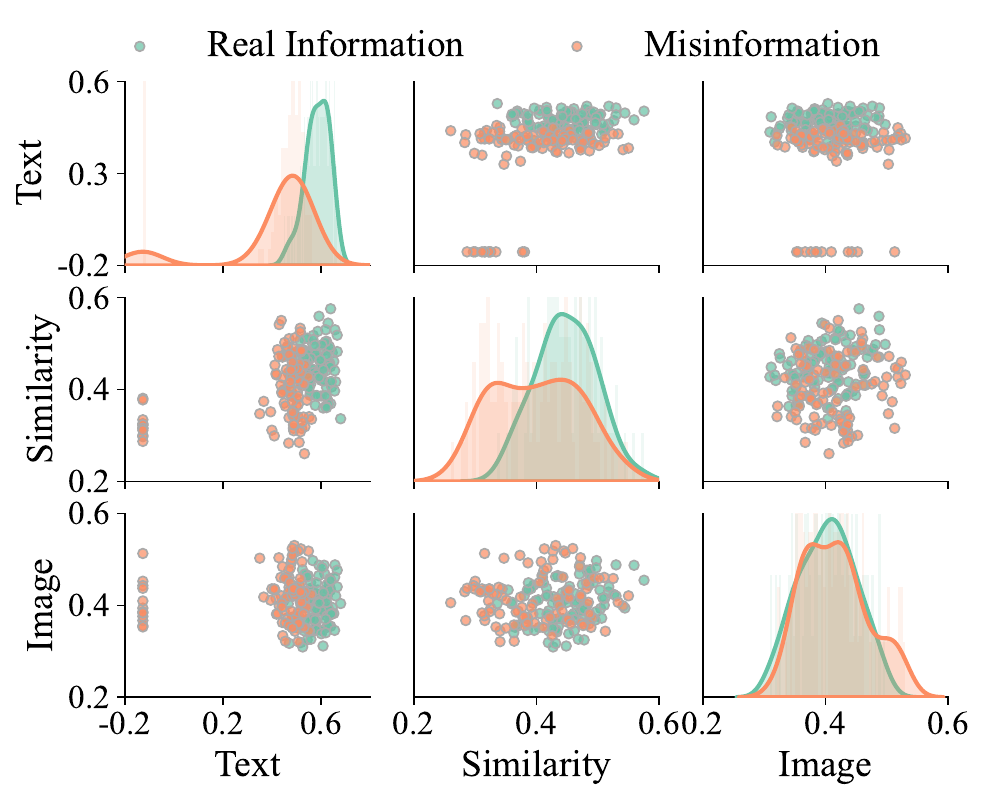}
    \caption{The joint distributions of three content consistency metrics for misinformation and real information. Misinformation and real information illustrate different distributions, especially in \textbf{Text} and \textbf{Similarity}.}
    \label{fig: correlation}
\end{figure}

\section{Basic Analysis of \ourdataset{}}
\label{sec: basic analysis}
\paragraph{\ourdataset{} is the most comprehensive.} We compare \ourdataset{} with the recent datasets for misinformation and social bots, illustrated in Table \ref{tab: summary}. \ourdataset{} is the only dataset simultaneously containing misinformation and social bot annotations. Meanwhile, from the misinformation perspective, \ourdataset{} contains the most complete multi-modal information, including textual content, user comments, repost messages, images, videos, and related users. \ourdataset{} is the largest and contains the richest visual modal data for misinformation.

\paragraph{Misinformation and real information in \ourdataset{} are distinguishable.} We aim to explore the role of social bots in amplifying misinformation spread, which requires misinformation and real information to be distinguishable. Thus, we analyze whether misinformation and real information are distinguishable from two perspectives: \emph{data distribution} and \emph{misinformation detector}.

From the \emph{data distribution} perspective, we first explore the differences in content consistency between misinformation and real information. We employ three metrics: (\romannumeral 1) \textbf{Text} to evaluate the text consistency of a specific instance and all instances; (\romannumeral 2) \textbf{Image} to evaluate the image consistency of a specific instance and all instances; and (\romannumeral 3) \textbf{Similarity} to evaluate the consistency of text and image in a specific instance. We provide the calculation formula in Appendix \ref{subapp: content agreement} and present the joint distributions in Figure \ref{fig: correlation}. It illustrates that misinformation and real information present distinct consistency. Specifically, real information presents higher \textbf{Text} and \textbf{Similarity}. Namely, we could conclude that misinformation and real information are distinguishable in terms of consistency. 

To further capture the image differences between misinformation and real information, we present the distribution of image categories and sentiments in Figure \ref{fig: category} in Appendix \ref{subapp: image distribution}. It illustrates that misinformation and real information present distinct distributions. Specifically, real information would contain more neutral images while misinformation would contain more screenshots. 

\begin{table}[t]
    \centering
    \resizebox{\linewidth}{!}{
    \begin{tabular}{l|c|c|c|c}
        \toprule[1.5pt]
         Models&Accuracy&F1-score&Precision&Recall\\
         \midrule[1pt]
         Vanilla&$95.2_{\pm 0.6}$&$92.3_{\pm 0.8}$&$93.7_{\pm 1.7}$&$91.0_{\pm 1.0}$\\
         \midrule[1pt]
         ~~w/o \emph{Interaction}&$\mathop{81.6^\star_{\pm4.5}}\limits_{14.2\%\downarrow}$&$\mathop{77.3^\star_{\pm4.1}}\limits_{16.2\%\downarrow}$&$\mathop{64.4^\star_{\pm5.9}}\limits_{31.3\%\downarrow}$&$\mathop{97.3^\star_{\pm 1.2}}\limits_{7.0\%\uparrow}$\\
         ~~w/o \emph{Vision}&$\mathop{94.1^\star_{\pm 0.5}}\limits_{1.1\%\downarrow}$&$\mathop{90.3^\star_{\pm1.0}}\limits_{2.2\%\downarrow}$&$\mathop{94.2^\dagger_{\pm1.2}}\limits_{0.4\%\uparrow}$&$\mathop{86.8^\star_{\pm2.1}}\limits_{4.6\%\downarrow}$\\
         ~~w/o \emph{Extra}&$\mathop{78.5^\star_{\pm5.2}}\limits_{17.5\%\downarrow}$&$\mathop{74.5^\star_{\pm 4.3}}\limits_{19.3\%\downarrow}$&$\mathop{60.6^\star_{\pm6.0}}\limits_{35.4\%\downarrow}$&$\mathop{97.3^\star_{\pm 1.0}}\limits_{6.9\%\uparrow}$\\
         \bottomrule[1.5pt]
    \end{tabular}
    }
    \caption{Performance of baseline and variants. We report the mean and standard deviation of ten-fold cross-validation. We also report the performance changes and conduct the paired t-test with vanilla, where $\star$ denotes the p-value is less than $0.0005$ and $\dagger$ denotes otherwise. Misinformation and real information are distinguishable with the help of user interactions.}
    \label{tab: main_results}
\end{table}

From the \emph{misinformation detector} perspective, we design a simple misinformation detector to verify whether the detector could identify misinformation in \ourdataset{}, where we provide the details of this model in Appendix \ref{subapp: misinformation detector}. We present the performance of the detector and ablation variants in Table \ref{tab: main_results}. This simple detector achieves remarkable performance, where the accuracy reaches 95.2\%. The ideal performance proves that misinformation and real information are easily distinguished by a machine detector, which helps explore the differences between social bots in spreading misinformation and real information. Meanwhile, the detector without \emph{interaction} drops to 77.3\% on f1-score, illustrating the effectiveness of user reactions, which coincides with our speculation that social bots might have different social patterns in misinformation and real information. We also provide a complete analysis of the ablation study in Appendix \ref{subapp: detector ablation study}.

\paragraph{\ourdataset{} has high social bot annotation quality, where the weakly supervised annotator is reliable.} The construction of the weakly supervised annotator contains three phases, where we have proven that each phase is reliable:
\begin{itemize}[topsep=4pt, leftmargin=*]
    \item \textbf{Preprocessing phase.} We recruited 315 human annotators, each of whom annotated 1,000 accounts and 20 standard accounts (the annotators did not know the standard accounts). Among them, 300 human annotators achieved more than 80\% accuracy on the standard accounts. The average accuracy of the reliable annotators on the standard accounts is 93.75\%. For the agreement between human annotators, the average Fleiss' Kappa is 0.4281, showing moderate agreement.
    \item \textbf{Training phase.} We employed multiple detectors aiming to identify various social bots. To ensure the annotator's credibility, we filtered in detectors achieving 80\% accuracy and obtained 4 detectors. The accuracy on the test set reaches 85.03\%, which is higher than TwiBot-20 \citep{feng2021twibot} and TwiBot-22 \citep{feng2022twibot}, illustrating credibility. We also provide the performance of each detector and the corresponding temperature in Appendix \ref{subapp: expert performance}.
    \item \textbf{Inference phase.} We randomly sample 50 social bots and 50 genuine accounts in \ourdataset{} and manually annotate them through a human expert. The Cohen's Kappa between the human expert and the automatic annotator is 0.74, showing good agreement.
\end{itemize}

\section{Misinformation and Social Bots}
\label{sec: analysis}

\paragraph{Social bots are deeply involved in information spread.} We first check the bot percentage:
\begin{itemize}[topsep=4pt, leftmargin=*]
    \item The whole \ourdataset{} contains 952,955 accounts, of which 411,635 are active. There are 65,749 social bots, accounting for 15.97\%.
    \item Among 5,750 accounts publishing misinformation, there are 3,799 active accounts. There are 767 social bots, accounting for 20.19\%.
    \item Among 226,235 accounts participating in the misinformation spread, 95,360 are active. There are 13,020 social bots, accounting for 13.65\%.
    \item Among 749,763 accounts participating in the real information spread, 325,414 are active. There are 54,253 social bots, accounting for 16.67\%.
\end{itemize}

\begin{figure}[t]
    \centering
    \includegraphics[width=\linewidth]{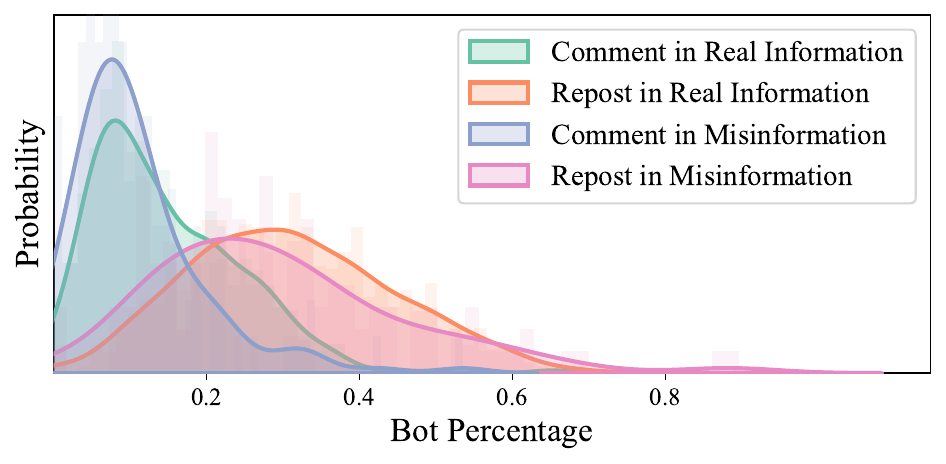}
    \caption{Probability density distributions of the percentage of social bots in information reposting and commenting. Social bots are deeply involved in information reposting and commenting.}
    \label{fig: bot_in_spread}
\end{figure}

Figure \ref{fig: bot_in_spread} further presents the distribution of social bots in information reposting and commenting. The average bot percentage of misinformation reposting and commenting is 29.3\% and 10.9\%, respectively, while the percentage of real information reposting and commenting is 31.1\% and 14.7\%. It illustrates that the distribution of misinformation and real information is similar, with slightly more social bots participating in spreading real information than misinformation. Meanwhile, reposting tends to have a higher bot percentage than commenting. Thus, we could conclude that social bots are deeply involved in information spread, where the main spread method is to repost information. 

\paragraph{Misinformation with the same topics has similar content, providing the basis of echo chambers, and social bots amplify this phenomenon.} We first group all pieces of misinformation into clusters with the same topics according to the \textbf{judgment}, where we provide the clustering algorithm in Appendix \ref{subapp: cluster algorithm}. We group 11,393 pieces of misinformation into 2,270 clusters, each of which represents a specific topic or event, \emph{e.g.}, ``The last two minutes of the air crash''. We aim to explore the textual content similarity of misinformation with the same topics and across different topics. 

\begin{figure}[t]
    \centering
    \includegraphics[width=\linewidth]{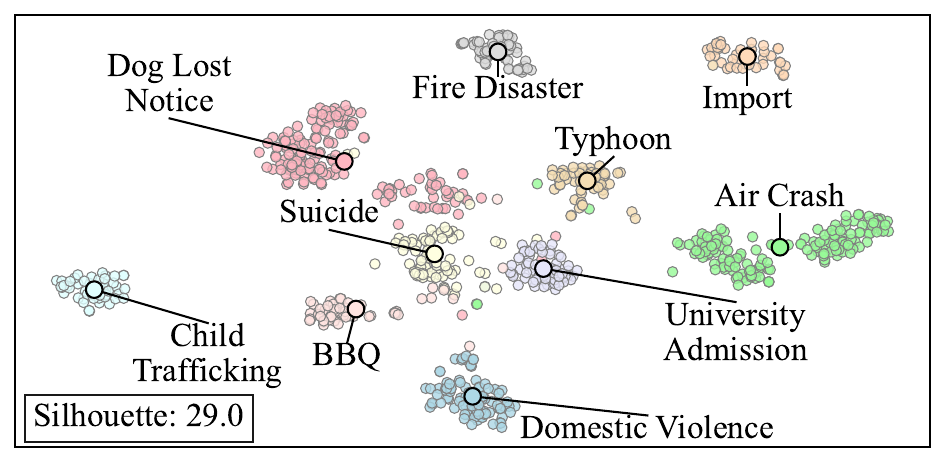}
    \caption{Visualization of misinformation content representations within the largest 10 clusters. Each dot corresponds to a misinformation instance colored according to its topic. The topic labels annotated by the \textbf{judgment} are plotted at each cluster center. We also calculate the silhouette score ($\times 100$). The cohesive clusters indicate misinformation about the same topic having similar content, providing the basis of echo chambers.}
    \label{fig: cluster}
\end{figure}

We first select the 10 largest clusters as representatives, since there is a long-tail effect in cluster size, where we present the selected clusters in Appendix \ref{subapp: top-ten topics}. We visualize the misinformation content representations in Figure \ref{fig: cluster}, which shows the BERT representation using t-SNE dimensionality reduction. It illustrates that the clusters are cohesive, where the silhouette score is 0.29. Namely, each cluster shares similar content while different clusters share significant differences. It suggests that the misinformation environment is homogeneous, providing the basis for echo chambers.

We conduct further quantitative analysis by calculating \emph{semantic-level} and \emph{token-level} pairwise scores between two instances, where higher scores mean the content of the two instances is more similar. For \emph{semantic level}, we employ the cosine similarity of the BERT representations, while for \emph{token level}, we leverage the ROUGE-L score, where we provide the detailed calculation in Appendix \ref{subapp: pairwise scores}. For \emph{semantic level}, the average value within the same cluster is 0.9448, and the others' average is 0.5847. For \emph{token level}, the average value within the same cluster is 0.7815, and the others' average is 0.0773. We also present completed values in Figure \ref{fig: heatmap} in Appendix \ref{subapp: score heatmap}. The quantitative results emphasize that misinformation with the same topics has similar content, and misinformation with different topics has distinct content.

\begin{figure}[t]
    \centering
    \includegraphics[width=\linewidth]{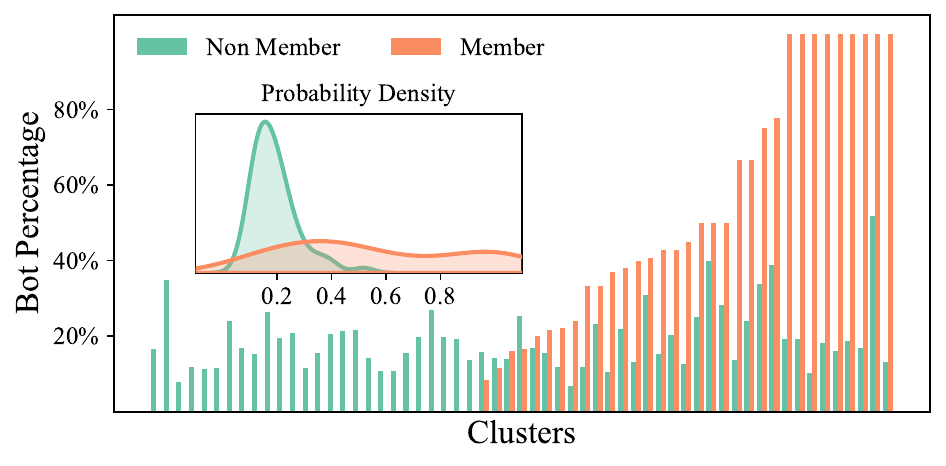}
    \caption{Bot percentage distribution comparison between echo chamber members and non-members across various clusters. Bot percentage among echo chamber members is generally higher than among non-members.}
    \label{fig: members}
\end{figure}

We finally explore the patterns of social bots in misinformation. We consider an account a potential \emph{echo chamber member} if it participates in at least two misinformation discussions (repost, comment, or like) in the same cluster. Figure \ref{fig: members} presents the distribution of bot percentage among echo chamber members and non-members within various clusters. It illustrates that around 18\% non-members are social bots. Meanwhile, the members do not contain bots in about half of the clusters. However, in the other half, members exhibit a higher bot percentage across most clusters compared to non-members, reaching up to 50\% in many clusters. We speculate that social bots engage in discussions involving misinformation on specific topics, thereby reinforcing the echo chamber effect.

\paragraph{Social bots generate similar content, aiming to manipulate public opinions.} Online information consumers are reluctant to process information deliberately \citep{moller2020explaining}, becoming susceptible to cognitive biases \citep{pennycook2018prior, vosoughi2018spread}. We aim to explore how public opinion changes and how social bots potentially manipulate it. We focus on how public sentiments and stances change in \ourdataset{}. We employ two existing classifiers to obtain the sentiments and stances since it is not our contribution, where we provide the details in Appendix \ref{subapp: sentiments and stances}.

\begin{itemize}[topsep=4pt, leftmargin=*]
    \item For \emph{public sentiments}, we categorize sentiments into \emph{neutral} and \emph{non-neutral} (including \emph{happy}, \emph{angry}, \emph{surprised}, \emph{sad}, and \emph{fearful}). Figure \ref{fig: sentiment} in Appendix \ref{subapp: sentiment distribution} presents sentiment distribution in different social texts. It illustrates that misinformation would publish more emotional content while real information would naturally be reported. On the other hand, public reactions are always emotional, where misinformation shows more anger while real information shows more happiness. Thus, public sentiments are emotional. We further explore the degree or extent to which public sentiments change over the information spread, introducing a variation measure:
    \begin{align*}
    v_{\Delta}=\sum_{k=1}^n|f(x_k)-f(x_{k-1})|,
    \end{align*}
    where $f(x_k)$ denotes neutral sentiment proportion at time $x_k$ and we provide the details of $x_k$ in Appendix \ref{subapp: sentiment variation}. Figure \ref{fig: variation} visualizes sentiment variation distribution, where a larger value means a more drastic change. The average values of misinformation and real information reach 0.287 and 0.225. It illustrates that public sentiment changes are dramatic during information spread.

    \begin{figure}[t]
    \centering
    \includegraphics[width=\linewidth]{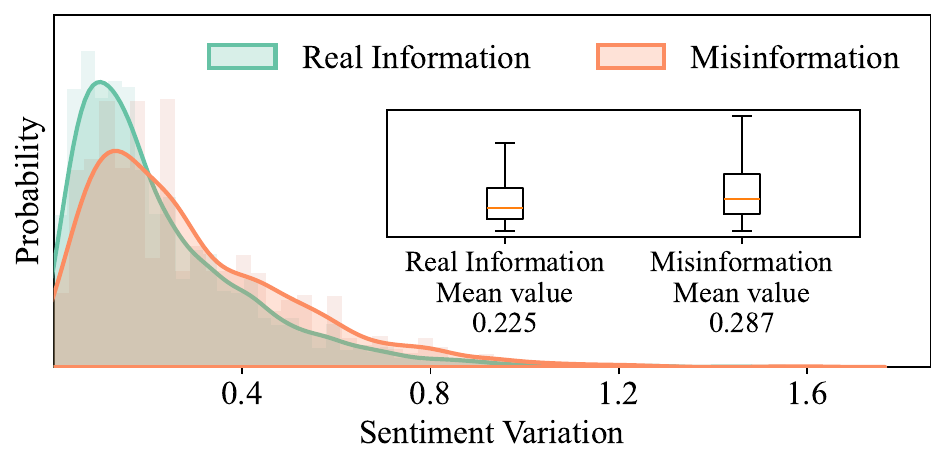}
    \caption{The distribution of neutral sentiment variation. Public sentiment changes are dramatic during information spread, with misinformation slightly more drastic.}
    \label{fig: variation}
\end{figure}

\begin{figure}[t]
    \centering
    \includegraphics[width=\linewidth]{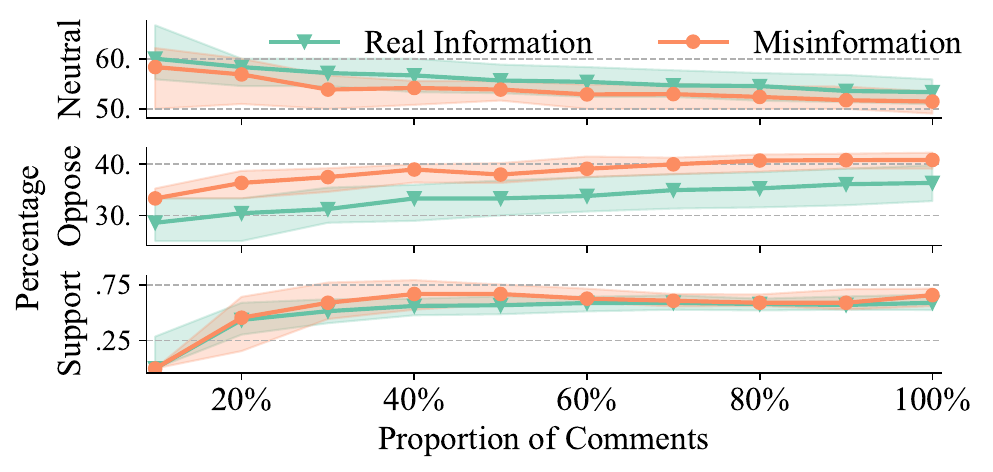}
    \caption{The proportion of comments with different stances as the comments increases. Public stances become increasingly polarized, where misinformation contains more comments with clear stances.}
    \label{fig: time}
\end{figure}

    \item For \emph{public stances}, we categorize stances into \emph{support}, \emph{oppose}, and \emph{neutral}. Figure \ref{fig: time} presents the proportion of each stance with the comments increasing over time. A striking finding is that only about 1\% accounts explicitly expressed a supportive stance, while the majority are neutral or opposed. Meanwhile, misinformation consistently presents higher opposition and lower neutrality. It illustrates that public stances become more polarized as the information spreads, where the neutral ratio suffers a drop of around 11\%.
\end{itemize}

Therefore, we can conclude that as the information spreads, public opinions, including sentiment and stance, become polarized, especially regarding misinformation. We then quantitatively prove the correlation between polarization and social bots by the Pearson correlation coefficient: the number of social bots demonstrates strong correlations with the number of comments with non-neutral stances (r = 0.6661) and sentiments (r = 0.6750). We also provide the completed coefficient in Appendix \ref{subapp: correlation coefficient}. The relatively high correlation coefficients indicate that social bots might influence public opinion.

\begin{figure}[t]
    \centering
    \includegraphics[width=\linewidth]{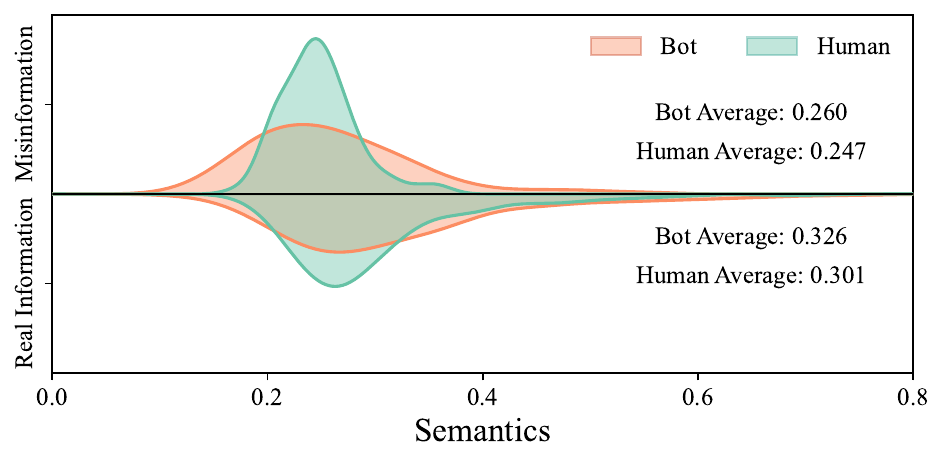}
    \caption{Distribution of social bots' and humans' semantic similarities, where social bots present higher similarities. Namely, social bots would publish similar content to manipulate public opinion.}
    \label{fig: echo_chamber}
\end{figure}

\begin{figure}[t]
    \centering
    \includegraphics[width=\linewidth]{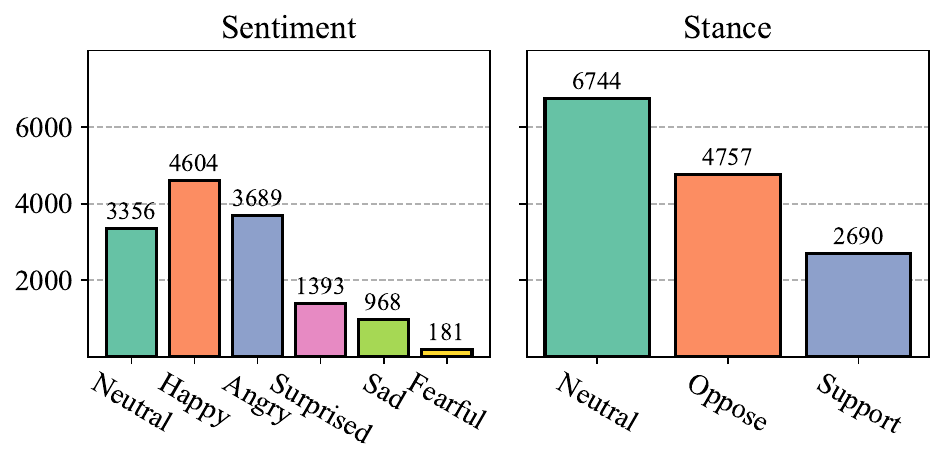}
    \caption{The sentiments and stances of comments published by social bots. Social bots would publish polarized content, manipulating public opinions.}
    \label{fig: bot_comment}
\end{figure}

We further explore social bot characteristics in information spread. We first calculate the semantic similarity of a specific account, where a higher value means that this account would publish more similar content. We present the detailed calculation method in Appendix \ref{subapp: semantic similarity} and present the results in Figure \ref{fig: echo_chamber}. It illustrates that social bots generally present higher values than humans. Social bots would publish similar content to amplify the \emph{bandwagon effect}, where online users adopt behaviors or actions simply because others are doing so, influencing the information spread \citep{wang2019jumping}. We then identify the sentiments and stances of social texts generated by social bots and present the results in Figure \ref{fig: bot_comment}. It demonstrates that social bots publish more emotional content and comments with clear stances. The results enhance the finding that social bots generate similar content, aiming to manipulate public opinions.

\section{Related Work}
\subsection{Misinformation Detection}
Mainstream detectors focuses on the information content, including text \citep{hartl2022applying, xiao2024msynfd, gong2024heterogeneous}, images \citep{liu2023interpretable, zhang2024escnet, zhang2024reinforced}, and videos \citep{tan2023deepfake, bu2024fakingrecipe, zeng2025imol}. They extract features such as emotion \citep{zhang2021mining, wan2025truth} and employ neural networks such as graph neural networks \citep{tao2024semantic, zhang2024heterogeneous, lu2024dual, wan2024fndpro} or neurosymbolic reasoning \citep{dong2024unveiling} to characterize information. 
Besides information content, the context, such as user interactions \citep{shu2019defend, lu2020gcan}, user profile \citep{sun2023hg, xu2024harnessing}, and evidence \citep{chen2024complex, wan-etal-2025-risk} provides helpful signals to detect misinformation. They would model propagation patterns \citep{cui2024propagation}, construct news environments \citep{yin2024gamc}, or extract multi-hop facts \citep{zhang2024causal} to enhance detection performance. Recently, to combat LLM-generated misinformation \citep{zhang2024llm, venkatraman2024gpt, wan2025difar}, models employing LLMs \citep{wan2024dell, nan2024let} through prompting \citep {guan2024language, hu2024bad} and in-context learning \citep{wang2024explainable} have been proposed.

\subsection{Social Bot Detection}
Social bot detectors fall into feature-, content-, and graph-based. 
Feature-based models conduct feature engineering for accounts \citep{feng2021satar, hays2023simplistic}. 
Content-based models employ NLP techniques \citep{lei2023bic, cai2024lmbot} to characterize the content. 
Graph-based models model user interactions as graph structures and employ graph neural networks \citep{feng2021botrgcn, yang2023fedack, zhou2023semi, liu2024segcn} in a semi-supervised way to identify bots. 
Many researchers are committed to exploring the risks and opportunities LLMs bring to bot detection \citep{DBLP:journals/debu/Tan023, feng2024does}. 
\subsection{Social Media Safety}
Social media safety has become more crucial \citep{mou2024unveiling}, where misinformation and social bots are two main factors harming online safety.  
Numerous datasets for misinformation \citep{li2024mcfend, qazi2024gpt, lin2024cfever, chencan} and social bots \citep{feng2021twibot, feng2022twibot, shi2023mgtab} are proposed. 
Based on these datasets, the generalization of detectors \citep{zhang2024t3rd, assenmacher2024you}, misinformation propagation pattern \citep{aghajari2023adopting, ashkinaze2024dynamics}, how to mitigate misinformation spread \citep{konstantinou2023nudging, su2024rumor, ghosh2024clock}, health-related misinformation \citep{yang2024fighting, shang2024domain}, source credibility \citep{carragher2024detection, mehta2024interactive}, user profiling \citep{morales2023geometry, zeng2024adversarial}, and bot communities \citep{liu2023botmoe, tan2023botpercent} are investigated. 
However, relatively little attention has been paid to the interplay between misinformation and social bots, thus, we bridge the gap in this paper.

\section{Conclusion}
In this paper, we proposed a novel dataset named \ourdataset{} containing information and annotations of misinformation and social bots. 
\ourdataset{} is the most comprehensive; misinformation and real information are distinguishable; and social bots have high annotation quality.
Extensive analysis illustrates that (\romannumeral 1) social bots are deeply involved in information spread; (\romannumeral 2) misinformation provides the basis of echo chambers, and social bots amplify this phenomenon; (\romannumeral 3) social bots generate similar content aiming to manipulate public opinions.

\section*{Acknowledgements}
This work is supported in part by the National Natural Science Foundation of China (No. 62192781, No. 62272374), the Natural Science Foundation of Shaanxi Province (No. 2024JC-JCQN-62), the Key Research and Development Project in Shaanxi Province (No. 2023GXLH-024), the Project of China Knowledge Center for Engineering Science and Technology, the Project of Chinese Academy of Engineering “The Online and Offline Mixed Educational Service System for ‘The Belt and Road’ Training in MOOC China”, and the K. C. Wong Education Foundation. This work was completed during the internship at SGIT AI Lab, State Grid Corporation of China.

\section*{Limitation}
Our proposed dataset is the largest containing misinformation and social bot annotations simultaneously. Meanwhile, it contains multiple modalities, including images and videos, and user interactions. However, due to the focus on news spread, it does not contain interactions like the friend relationship, missing potential relations between social bots and genuine accounts. Meanwhile, we propose a weakly supervised framework for annotating social bots, whose accuracy is comparable to that of crowd-sourcing. However, it struggles to achieve better recall and might miss several social bots. Finally, the experiments in this work focus primarily on the Sina Weibo platform. We expect to expand our analysis to other social media platforms such as $\mathbb{X}$ (Twitter) or Reddit, in future work. 

\section*{Ethics Statement}
Research on misinformation and social bots is essential for countering online malicious content. This research demonstrates that social bots would amplify the spread of misinformation, enhancing echo chambers and manipulating public opinions. However, it may increase the risk of dual-use, where malicious actors may develop advanced social bots to spread misinformation. We will establish controlled access to ensure that the trained annotator checkpoints are only publicly available to researchers. Meanwhile, we will hide the privacy information in the dataset when we publish it.

Our models are trained on crowd-sourced data, which might contain social biases, stereotypes, and spurious correlations. Thus, our model would provide incorrect annotations. We argue that the predictions of our models should be interpreted as an initial screening, while content moderation decisions should be made with experts in the loop.

\bibliography{custom}

\appendix

\section{Details of \ourdataset{} Dataset} 

\begin{figure*}
    \centering
    \includegraphics[width=\linewidth]{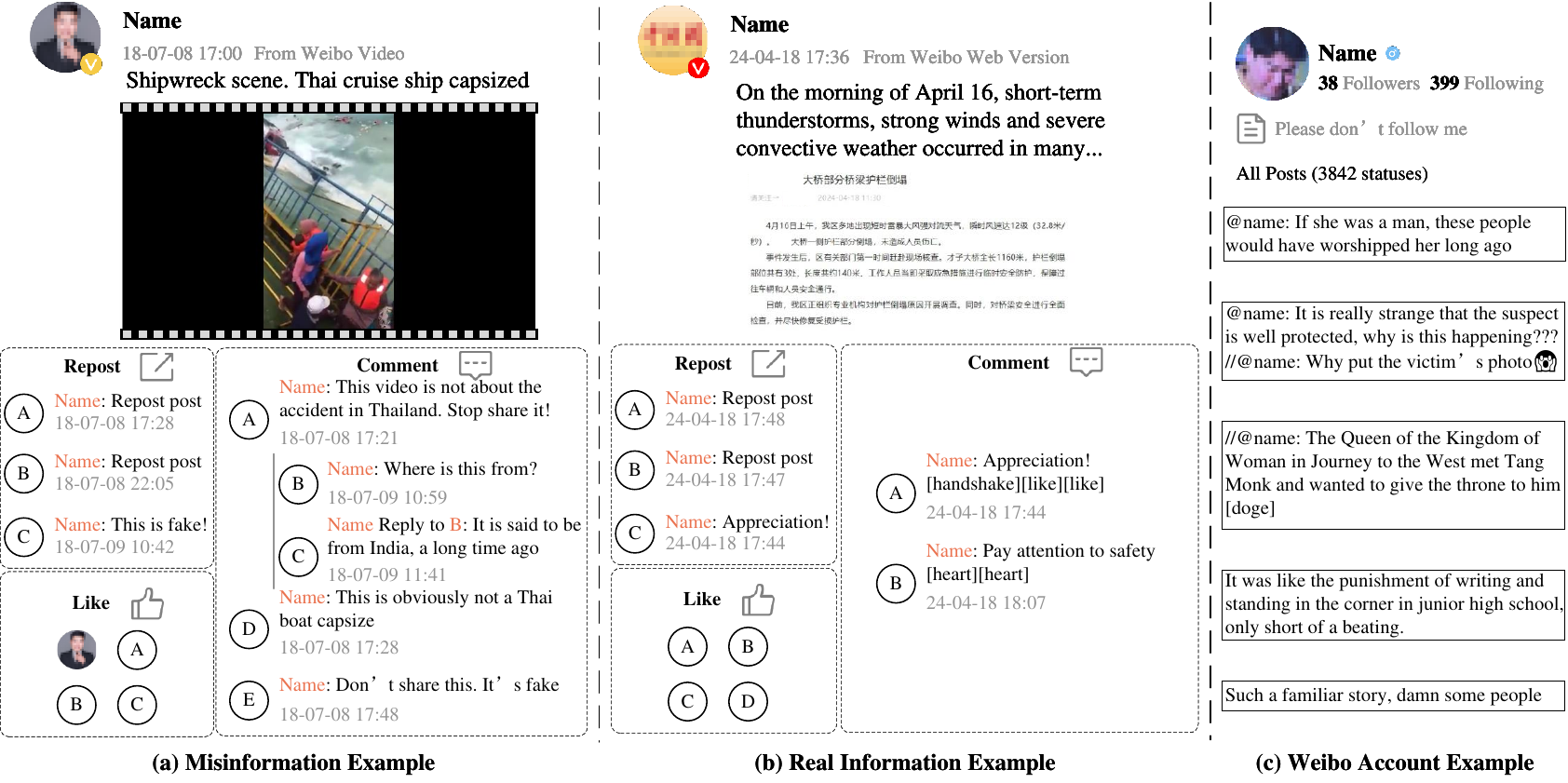}
    \caption{The examples in \ourdataset{}. We present (a) a misinformation example, (b) a real information example, and (c) a Weibo account example. We translate original information into English and conceal the private information.}
    \label{fig: example}
\end{figure*}

\subsection{Examples in \ourdataset{}}
\label{subapp: dataset example}
Formally, an online information instance is represented as $A=\{s, I, V, \mathcal{G}_{\textit{repost}}, \mathcal{G}_{\textit{comment}},\mathcal{U},y\}$. The image set $I=\{I_i\}$ contains multiple images while the video set $V=\{V_i\}$ contains multiple videos. The repost graph $\mathcal{G}_{\textit{repost}}=\{\mathcal{V}, \mathcal{E}, \mathcal{T}\}$ is a dynamic text-attributed graph (or tree) where the center node is the information content and another node $v\in\mathcal{V}$ denotes a repost text, $e=(v_i, v_j)\in\mathcal{E}$ denotes a repost relation connecting $v_i$ and $v_j$, and $\mathcal{T}:\mathcal{V}\rightarrow\mathbb{R}$ denotes the published time of each node. $\mathcal{G}_{\textit{comment}}=\{\mathcal{G}_\textit{comment}^i\}$ denotes the comment graph set, where each comment graph $\mathcal{G}_\textit{comment}^i$ is a dynamic text-attributed graph (or tree). Each comment graph is similar to the repost graph except for the center node, where the center node is a comment that directly comments on the information. Besides, a Weibo account instance is represented as $U=\{F, T, y\}$. The feature set contains \emph{follower count}, \emph{following count}, \emph{status count}, \emph{verified} (2 types), \emph{svip} (2 types), \emph{account type} (10 types), and \emph{svip level} (6 types). The post set $T$ contains the most recent five posts in the user timeline. We provide a piece of misinformation, real information, and a Weibo account example in Figure \ref{fig: example}.

\begin{figure}
    \centering
    \includegraphics[width=\linewidth]{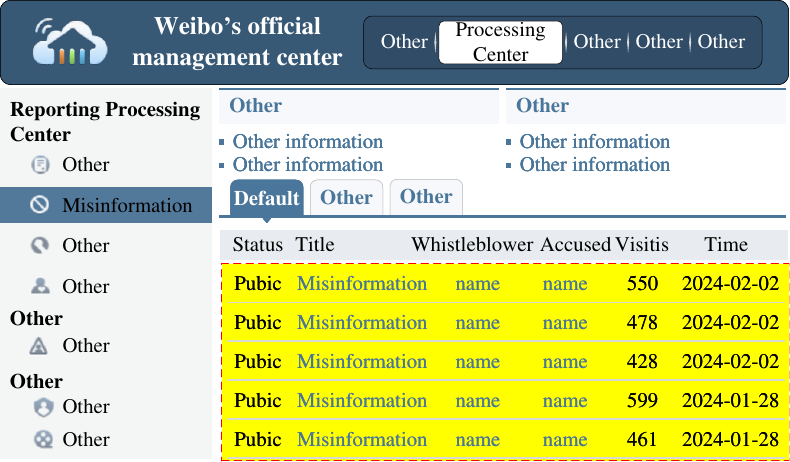}
    \caption{The overview of the Weibo's official management center. We conceal private or unrelated information and translate the main information into English. We highlight the misinformation items.}
    \label{fig: platform}
\end{figure}

\subsection{Management Center}
\label{subapp: management center}
The Weibo's official management center is a Weibo official. Here is the link: \url{https://service.account.weibo.com/?type=5&status=0}. If the users are logging into the platform for the first time, it will redirect to the Weibo homepage (\url{https://weibo.com/}). After logging in with a Weibo account, entering the platform again will lead to the right platform homepage. Figure \ref{fig: platform} shows the overview of this platform, where we conceal private or unrelated information and translate the main language into English. If the users successfully log into this platform, they will view a similar website. It is worth noting that the number of instances that each logged-in user can access per day is limited, so it took us about 10 months to collect all the misinformation on the platform.

\begin{table}[t]
    \centering
    \resizebox{\linewidth}{!}{
    \begin{tabular}{p{0.95\linewidth}}
    \toprule[1.5pt]
    \textbf{Post flagged as misinformation}: Recently, in xxx, a "naughty child" took scissors and cut off the hair of a female customer in a barber shop when no one was paying attention. After the female customer called the police and negotiated, the parents compensated 11,500 yuan.\\
    \midrule[1pt]
    \textbf{Judgment}: After investigation, it was found that the Weibo post claiming that "a woman's hair was cut off by a naughty child and her parents paid her 10,000 yuan in compensation" actually happened in May 2023, not recently. The respondent's speech is "outdated information" and constitutes "publishing false information"\\
    \bottomrule[1.5pt]
    \end{tabular}
    }
    \caption{An example of misinformation and corresponding \textbf{judgment} (translated into English). The \textbf{judgment} provides a basis for identifying misinformation topics.}
    \label{tab: example}
\end{table}

\subsection{Misinformation Example}
\label{subapp: misinformation example}
After logging in to the platform, it mainly contains users' posts flagged as misinformation and a corresponding \textbf{judgment}. The platform moderators or police flag the misinformation and publish the \textbf{judgment}. We provide an example in Table \ref{tab: example}. The judgment is the same for different pieces of misinformation on the same event. 
\subsection{Entity Filter}
\label{subapp: entity filter}
We obtained 7,445 entities using the keyphrase extractor. We employ two strategies to filter out noisy entities:
\begin{itemize}[topsep=4pt, leftmargin=*]
    \item Frequency less than 10. These entities appear occasionally in misinformation and are unlikely to cause entity bias. We only focus on common entities that appear in large numbers in misinformation, so we need to ensure that they appear at a similar frequency in real information. 
    \item The number of characters is 1. These entities might come from the noises of the keyphrase detector. Meanwhile, these entities may not contain enough semantic information.
\end{itemize}
After filtering, we obtained 1,961 entities. We believe these entities are common and contain rich semantic information. As a result, it would mitigate the effects of entity bias if real information also contains these entities.

\subsection{Query Method}
\label{subapp: query method}
\begin{figure}[t]
    \centering
    \includegraphics[width=\linewidth]{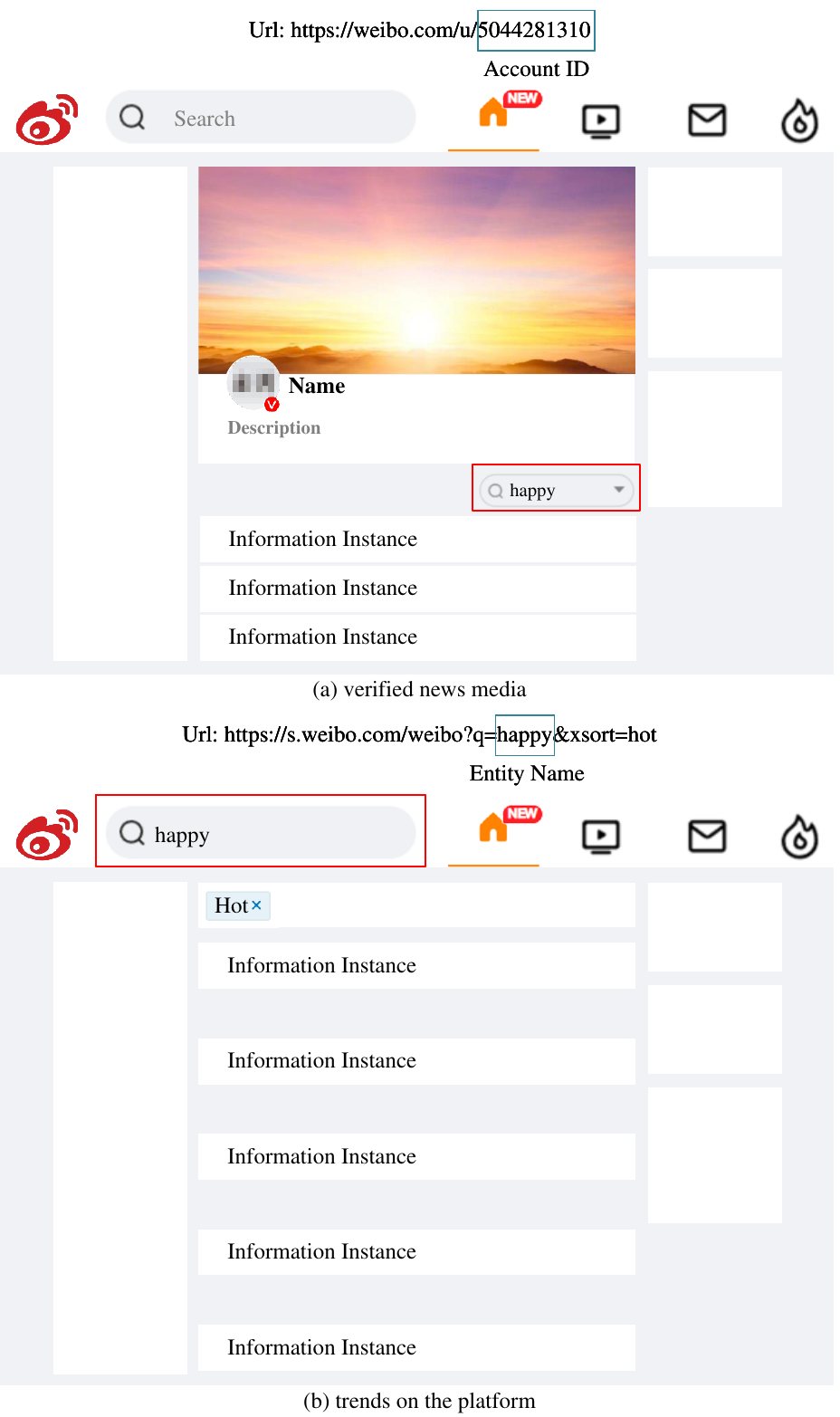}
    \caption{The overview of the two search pages. The red box presents the search functions.}
    \label{fig: search}
\end{figure}

We mainly employ the official search function of the Weibo platform to search the given entity. Given an entity, the search function will return several posts containing the entity. 
\begin{itemize}[topsep=4pt, leftmargin=*]
    \item \textbf{Verified news media}. After entering a specific account's homepage, we could use the search function to search posts in this account.
    \item \textbf{Trends on the platform}. Given an entity, such as \emph{happy}, we collect posts in the trends using \url{https://s.weibo.com/weibo?q=happy&xsort=hot}.
\end{itemize}
Figure \ref{fig: search} presents an overview of these two search functions, where the red box illustrates the search function.

\subsection{Verified Accounts}
\label{subapp: verified accounts}
We employ 8 verified news accounts, and Table \ref{tab: verified accounts} presents the information about them. They have a red ``verified'' symbol. When an account has more than 10,000 followers and this account has been read more than 10 million times in the last 30 days, it can obtain the red ``verified'' symbol. 

\begin{table}[t]
    \centering
    \resizebox{\linewidth}{!}{
    \begin{tabular}{c|c|c|c}
        \toprule[1.5pt]
         Homepage&Follower Count&Status Count&Discussion Count\\
         \midrule[1pt]
         \url{https://weibo.com/u/1496814565}&33.8 million&225.3 thousand&334.0 million\\
         \url{https://weibo.com/u/5044281310}&32.6 million&163.2 thousand&573.0 million\\
         \url{https://weibo.com/u/1618051664}&111.0 million&302.6 thousand&1.6 billion\\
         \url{https://weibo.com/u/1974808274}&3.3 million&58.8 thousand&27.3 million\\
         \url{https://weibo.com/u/2028810631}&107.0 million&166.4 thousand&469.0 million\\
         \url{https://weibo.com/u/2656274875}&137.0 million&187.8 thousand&3.7 billion\\
         \url{https://weibo.com/u/1784473157}&81.5 million&246.5 thousand&786.0 million\\
         \url{https://weibo.com/u/1642512402}&62.4 million&224.4 thousand&410.0 million\\
         \bottomrule[1.5pt]
    \end{tabular}
    }
    \caption{The information about the selected verified news accounts. We provide the homepage links of them. They have a huge number of followers and discussions. }
    \label{tab: verified accounts}
\end{table}

\subsection{Source Credibility}
\label{subapp: source credibility}
Here we discuss the credibility of the two real information sources:
\begin{itemize}[topsep=4pt, leftmargin=*]
    \item \textbf{Verified news media}. These accounts are operated by legitimate news media and verified by the Weibo platform. Thus we believe this source is credible. 
    \item \textbf{Trends on the platform}. Weibo is a responsible social platform, where content moderators are efficient. As a result, the content moderation mechanism makes it easier to moderate posts with a lot of discussion. Because users would report the posts that they think are fake. After receiving reports, moderators only need to verify the post content instead of the whole discussion. It takes only a few days to moderate misinformation on the training. Meanwhile, the posts we collected are from one month ago in the trend. There is plenty of time for moderation.
\end{itemize}
\subsection{Annotation Guideline}
\label{subapp: annotation guideline}
We first summarize the general criteria to identify a social bot on Weibo: (\romannumeral 1) reposting or publishing numerous advertisements, (\romannumeral 2) devoted fans of a star publishing numerous related content, (\romannumeral 3) containing numerous reposting content without pertinence and originality, (\romannumeral 4) publishing numerous unverified and negative information, (\romannumeral 5) containing numerous posts with the ``automatically'' flags, (\romannumeral 6) repeated posts with the same content, and (\romannumeral 7) containing content that violates relevant laws and regulations. 

\begin{figure*}[t]
    \centering
    \includegraphics[width=\linewidth]{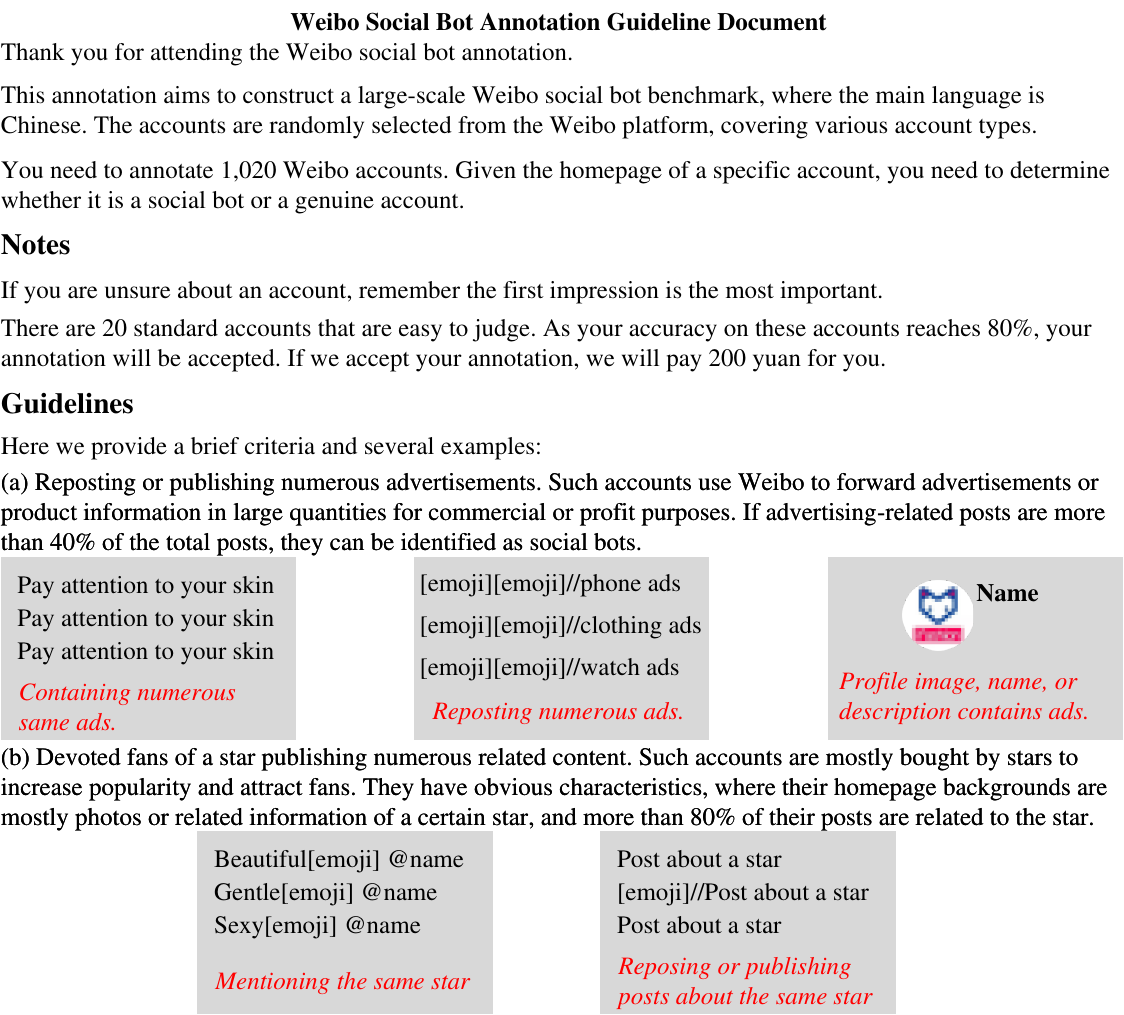}
    \caption{The overview of the guideline document, where we translate it into English. The human annotators are required to read this document before annotation. }
    \label{fig: document}
\end{figure*}

\begin{figure*}[t]
    \centering
    \includegraphics[width=\linewidth]{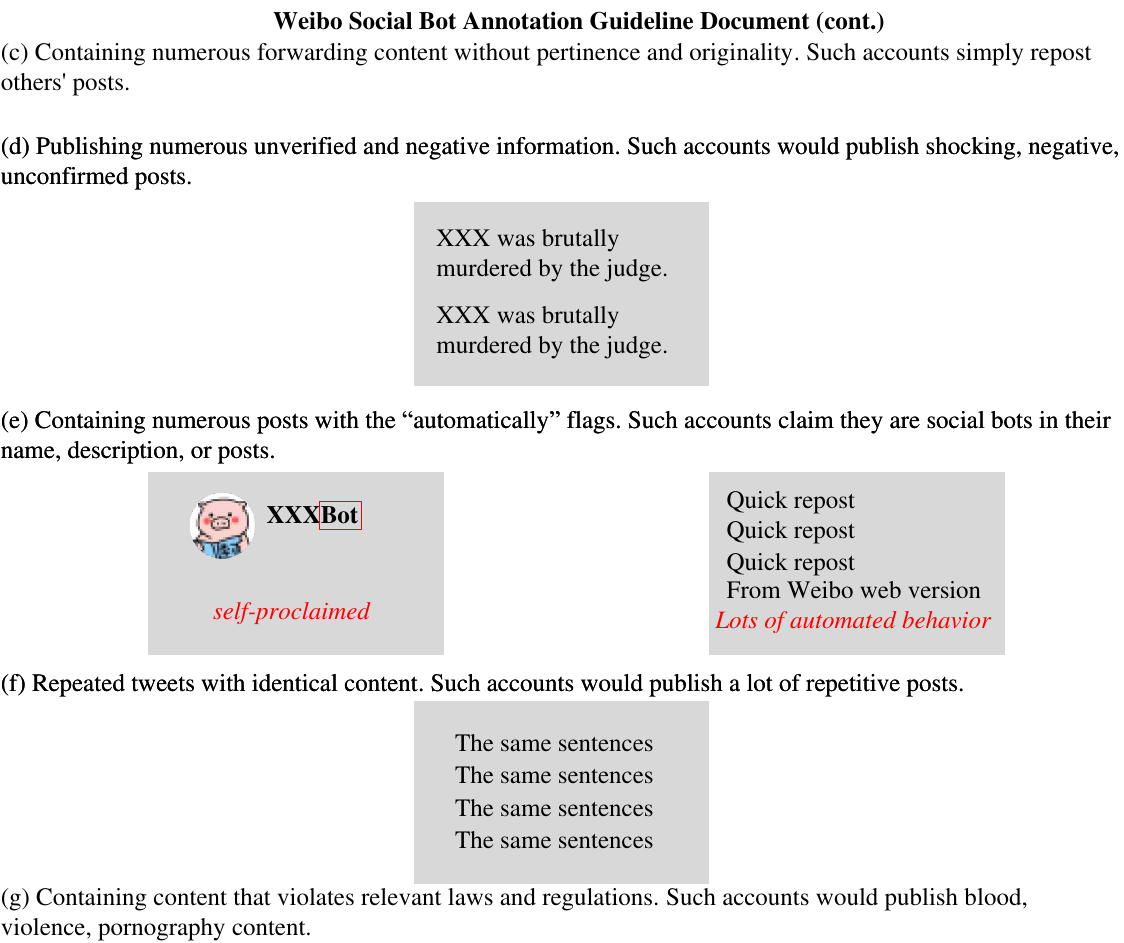}
    \caption{The overview of the guideline document (cont.).}
    \label{fig: document 2}
\end{figure*}

Based on the criteria, we write a guideline document for human annotators in Figures \ref{fig: document} and \ref{fig: document 2}. Each human annotator must read this document before annotating.

\subsection{Annotation Cost}
\label{subapp: annotation cost}
Each human annotator is required to annotate 1,000 accounts plus 20 standard accounts. If a human annotator achieves more than 80\% accuracy on the standard accounts, we will adopt the annotator's annotation. We will pay 200 yuan (about 28 dollars) for each qualified annotator. We recruited 315 annotators and, 300 are qualified. The crowd-sourcing takes about 60 days and costs 60,000 yuan. 
\subsection{Active Accounts}
\label{subapp: active accounts}
We focus on the active accounts in \ourdataset{} and this paper. According to the human annotators' feedback and the characteristics of the Weibo platform. If an account publishes more than five posts with a length of no less than five characters in the timeline, then we consider this account active. 
\subsection{Expert Settings}
\label{subapp: expert settings}
In the training phase, we leverage three categories of social bot detectors as experts:
\paragraph{Feature-based Detectors} We first preprocess the selected initial features to obtain the features for classifiers. For the \textbf{numerical} features (including \emph{follower count}, \emph{following count}, and \emph{status count}), we employ z-score normalization:
\begin{align*}
    z=\frac{x-\mu}{\sigma},
\end{align*}
where $x$ is the initial feature, $z$ is the preprocessed feature, and $\mu$ and $\sigma$ are the average and standard deviation in the training set. The average values are $5074.88$, $420.59$, and $1432.10$, while the standard deviation values are $283145.61$, $584.40$, and $1373.91$. For the \textbf{categorical} features (including \emph{verified}, \emph{svip}, \emph{account type}, and \emph{svip level}), we employ one-hot to obtain the initial representations. After that, we concatenate numerical and categorical representations to obtain the account representation $\boldsymbol{x}_\textit{f}$. After that, We employ MLP layers, random forests, and Adaboost as detectors. We adopt three feature-based experts (three classic classifiers).
\paragraph{Content-based Detectors} We employ \emph{name}, \emph{description}, and \emph{posts} to identify social bots. We assuming the notation of \emph{name} is $\boldsymbol{s}_{\textit{name}}$, of \emph{description} is $\boldsymbol{s}_{\textit{desc}}$, and of \emph{posts} is $\{\boldsymbol{s}_{\textit{post}}^i\}_{i=1}^N$ (here are $N$ posts). Given a text $\boldsymbol{s}$, we employ encoder-based language model to obtain the representation:
\begin{align*}
    \boldsymbol{x}=\mathrm{LM(\boldsymbol{s})}.
\end{align*}
For \emph{posts}, we average the representation:
\begin{align*}
    \boldsymbol{x}_{\textit{post}}=\frac{1}{N}\boldsymbol{x}^i_\textit{post}.
\end{align*}
We feed the representations into an MLP layer to identify social bots. We employ the pre-trained parameters of the encoder-based language models and do not update the parameters. We employ the parameters in the Hugging Face for BERT\footnote{Here is the \href{https://huggingface.co/google-bert/bert-base-chinese}{model link}.} and DeBERTa\footnote{Here is the \href{https://huggingface.co/IDEA-CCNL/Erlangshen-DeBERTa-v2-97M-Chinese}{model link}.}. We adopt six content-based experts (two encoder-based language models and three categories of texts).

\paragraph{Ensemble Detectors} We first employ MLP layers to transfer the feature-based and content-based representations and concatenate them:
\begin{align*}
    \boldsymbol{x}=\|_{i\in\{\textit{f},\textit{name},\textit{desc},\textit{post}\}}\mathrm{MLP}(\boldsymbol{x}_i).
\end{align*}
We adopt two ensemble experts (two encoder-based language models).

For all experts, we do not update the language model parameters. We set the \emph{hidden dim} as 256, \emph{learning rate} as $10^{-4}$, \emph{weight decay} as $10^{-5}$, \emph{batch size} as 64, \emph{dropout} as 0.5, \emph{optimizer} as Adam, \emph{activation function} as LeakyReLU.

We do not employ graph-based detectors because neighbor information is hard to access on the Weibo platform and would cost a lot during the inference process. Besides, the automatic annotator already achieves acceptable annotation quality.

\subsection{Temperature Settings}
\label{subapp: temperature settings}
Temperature scaling is a post-precessing technique to make neural networks calibrated. It divides the logits (the output of the MLP layers and the input to the softmax function) by a learned scalar parameter,
\begin{align*}
    p_i=\frac{e^{z_i/\tau}}{\sum_{j\in\mathcal{Y}}e^{z_j/\tau}},
\end{align*}
where $\mathcal{Y}$ denotes the label set, $p_i$ is the probability of belonging to category $i$. We learn the temperature parameter $\tau$ on the validation set. We conduct a grid search from 0.5 to 1.5 with an interval of 0.001, obtaining the optimal value by minimizing the expected calibration error on the validation set. 

\section{Details of Basic Analysis}
\subsection{Content Agreement Metrics}
\label{subapp: content agreement}
These three metrics are proposed to calculate the multi-modal content consistency, where a higher value means higher consistency. Formally, assuming each information instance is presented as $(T_i, I_i)$ (here we only focus on the textual and image content). Meanwhile, the information set is represented as $\{(T_i, I_i)\}_{i=1}^N$ (misinformation set or real information set). Given an instance $(T_i, I_i)$, we calculate \textbf{Text} as:
\begin{align*}
    \textit{text}_i=\frac{1}{N}\sum_{j=1}^N\mathrm{cosine}(\mathrm{BERT}(T_i),\mathrm{BERT}(T_j)),
\end{align*}
where $\mathrm{cosine}(\cdot)$ denote the cosine similarity function, $\mathrm{BERT}(\cdot)$ denote the BERT encoder\footnote{Here is the \href{https://huggingface.co/shibing624/text2vec-base-chinese}{model link}.}. We calculate \textbf{Similarity} as:
\begin{align*}
    \textit{similarity}_i=\mathrm{cosine}(\mathrm{CLIP}(T_i), \mathrm{CLIP}(I_i)),
\end{align*}
where $\mathrm{CLIP}(\cdot)$ denote the CLIP encoder\footnote{Here is the \href{https://huggingface.co/OFA-Sys/chinese-clip-vit-base-patch16}{model link}.}. We calculate \textbf{Image} as:
\begin{align*}
    \textit{image}_i=\frac{1}{N}\sum_{j=1}^N\mathrm{cosine}(\mathrm{ViT}(I_i),\mathrm{ViT}(I_j)),
\end{align*}
where $\mathrm{ViT}(\cdot)$ denote the ViT encoder\footnote{Here is the \href{https://huggingface.co/google/vit-base-patch16-224-in21k}{model link}.}.

\subsection{Image Distribution}
\label{subapp: image distribution}
\begin{figure}[t]
    \centering
    \includegraphics[width=\linewidth]{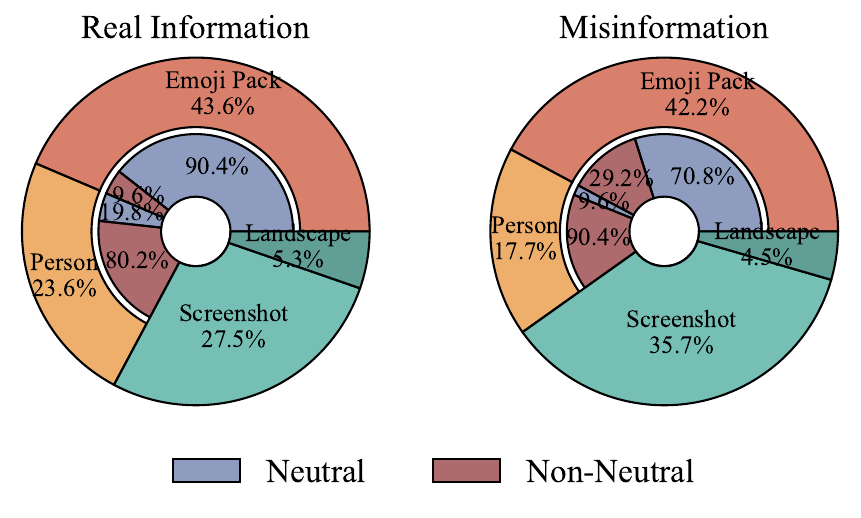}
    \caption{Image distribution of misinformation and real information, including categories and sentiments. Misinformation presents a different distribution from real information.}
    \label{fig: category}
\end{figure}

To further explore the differences in image distribution between misinformation and real information, we check the categories of the image in information. We select four common categories: (\romannumeral 1) \emph{person}, (\romannumeral 2) \emph{emoji pack}, (\romannumeral 3) \emph{landscape}, and (\romannumeral 4) \emph{screenshot}. We then investigate the sentiments of \emph{person} and \emph{emoji pack}, where \emph{person} is realistic and \emph{emoji pack} is virtual. The sentiments include neutral and non-neutral (angry, surprised, fearful, sad, and happy). To obtain the categories and sentiments, we employ pre-trained CLIP \citep{radford2021learning}\footnote{Here is the \href{https://huggingface.co/openai/clip-vit-base-patch16}{model link}.} in zero-shot format. Figure \ref{fig: category} presents the image distribution of misinformation and real information. Images in real information tend to focus on people, while misinformation prefers to publish screenshots. Regarding sentiment, most of the images related to people in both real and misinformation are non-neutral, proving that information publishers tend to employ appealing pictures. For virtual images, emoji packs in real information are predominantly neutral, with a small partial being non-neutral. However, most emoji packs in misinformation are still neutral, significantly less than those in real posts. Furthermore, we analyze the correlation between the sentiment of images and text content (Appendix \ref{subapp: sentiments and stances}), where 78.2\% of real information contains images with the same sentiment as the text while only 34.1\% of misinformation does. It further proves that misinformation and real information in \ourdataset{} are distinguishable. 

\subsection{Misinformation Detector}
\label{subapp: misinformation detector}
\begin{figure}[t]
    \centering
    \includegraphics[width=\linewidth]{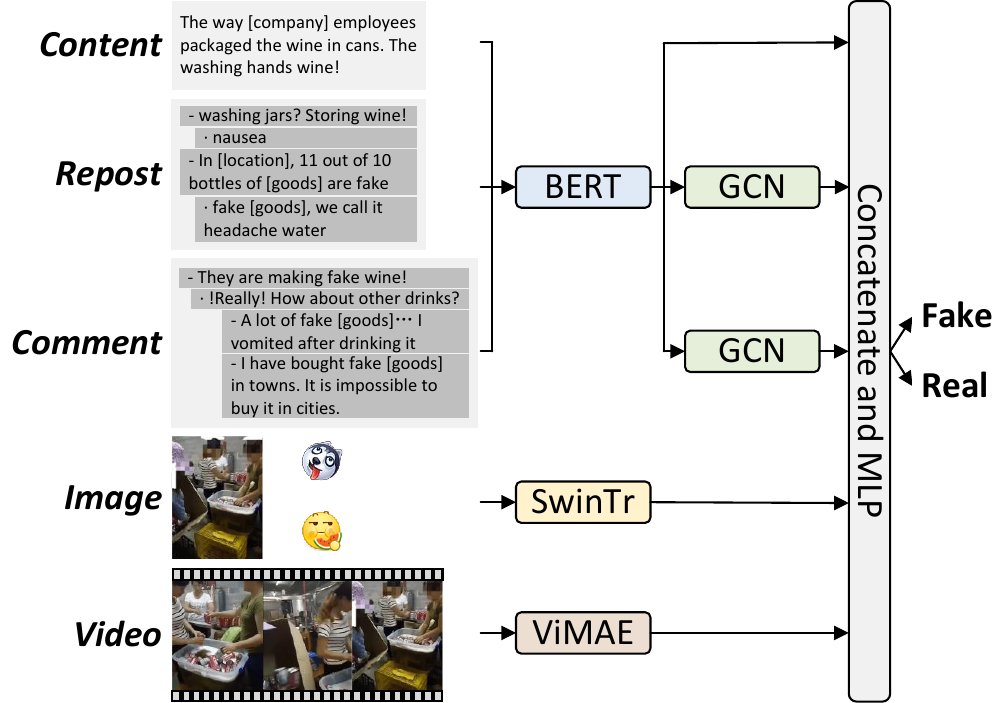}
    \caption{Overview of the misinformation detector, which employs multiple modality encoders to encode variance modalities and employs an MLP layer to identify misinformation.}
    \label{fig: model}
\end{figure}
We propose a simple misinformation detector as Figure \ref{fig: model} illustrates. We employ multi-modal encoders to encode \emph{content}, \emph{repost}, \emph{comment}, \emph{image}, and \emph{video}. For \emph{content}, we employ an encoder-based language model $\mathrm{LM}(\cdot)$\footnote{Here is the \href{https://huggingface.co/google-bert/bert-base-chinese}{model link}.} to encode content:
\begin{align*}
    \boldsymbol{f}_{\textit{content}} = \mathrm{LM}(s).
\end{align*}
To encode \emph{repost}, we employ the same language model $\mathrm{LM}(\cdot)$ to encode text-attributed node $v_i$ and obtain $\boldsymbol{h}_{v_i}^{(0)}$. We employ $L$ graph neural network layers to make each node interact:
\begin{align*}
    \boldsymbol{h}_{v_i}^{(\ell)}=\mathop{\mathrm{Aggr}}\limits_{\forall v_j\in \mathcal{N}(v_i)}(\{\mathrm{Prop}(\boldsymbol{h}_{v_i}^{(\ell-1)};\boldsymbol{h}_{v_j}^{(\ell-1)})\}),
\end{align*}
where $\mathcal{N}(v_i)$ denotes the set of neighbors of node $v_i$, $\mathrm{Aggr}(\cdot)$ and $\mathrm{Prop}(\cdot)$ are aggregation and propagation functions, where GCN \citep{DBLP:conf/iclr/KipfW17} is employed in practice. we finally employ the mean pooling operator as the $\mathrm{Readout}(\cdot)$ function to obtain the graph-level representation:
\begin{align*}
    \boldsymbol{f}_{\textit{repost}}=\mathrm{Readout}(\{\boldsymbol{h}_{v_i}^{(\ell)}\}_{v_i\in\mathcal{V}}).
\end{align*}
To encode \emph{comment}, we employ the same encoding method as \emph{repost} to obtain the representation of each comment graph $\mathcal{G}_{\textit{comment}}^i$ \citep{yang2023wsdms}. We then consider the average representations as the final representation:
\begin{align*}
    \boldsymbol{f}_{\textit{comment}}=\frac{1}{m}\boldsymbol{f}_{\textit{comment}}^i,
\end{align*}
where $m$ is the number of comment graphs. To encode \emph{image}, we employ a pre-trained swin transformer\footnote{Here is the \href{https://huggingface.co/microsoft/swinv2-tiny-patch4-window16-256}{model link}.} \citep{liu2022swin} $\mathrm{SwinTr}(\cdot)$ to obtain the representations of each image and adopt mean pooling to obtain the final representation:
\begin{align*}
    \boldsymbol{f}_{\textit{image}}=\mathrm{mean}(\mathrm{SwinTr}(I_i)),
\end{align*}
where $\mathrm{mean}(\cdot)$ denotes the meaning operator. To encode \emph{video}, we sample 256 frames from each video and resize each frame into $224\times 224$. We employ pre-trained VideoMAE\footnote{Here is the \href{https://huggingface.co/MCG-NJU/videomae-base}{model link}.} \citep{tong2022videomae} $\mathrm{VideoMAE}(\cdot)$ . For each time step, we take 16 frames and set the interval to 12 frames. We could obtain:
\begin{align*}
    \boldsymbol{f}_{\textit{video}}=\mathrm{mean}(\mathrm{VideoMAE}(V_i)).
\end{align*}
Finally, we concatenate them to obtain the representation of each user post:
\begin{align*}
    \boldsymbol{f}=[\boldsymbol{f}_{\textit{content}}\|\boldsymbol{f}_{\textit{repost}}\|\boldsymbol{f}_{\textit{content}}\|\boldsymbol{f}_{\textit{image}}\|\boldsymbol{f}_{\textit{video}}].
\end{align*}

Given an information instance $A$ and corresponding label $y$, we calculate the probability of $y$ being the correct prediction as $p(y\mid A)\propto \exp(\mathrm{MLP}(\boldsymbol{f}))$, where $\mathrm{MLP}(\cdot)$ denote an MLP classifier. We optimize this model using the cross-entropy loss and predict the most plausible label as $\arg \max_{y}p(y \mid A)$. The hyperparameter settings of the baseline are presented in Table \ref{tab: hyper} to facilitate reproduction. We conduct ten-fold cross-validation to obtain a more robust conclusion. When split folds, we do not split misinformation from the same topic (Appendix \ref{subapp: cluster algorithm}) into two folds to avoid data leakage. 

\begin{table}[t]
    \centering
    \resizebox{\linewidth}{!}{
    \begin{tabular}{c|c|c|c}
        \toprule[1.5pt]
         \textbf{Hyperparameter}&\textbf{Value}&\textbf{Hyperparameter}&\textbf{Value}\\
         \midrule[1pt]
         BERT embedding dim&768&optimizer&Adam\\
         GNN layers&2&learning rate&10$^{-4}$\\
         GNN embedding dim&256&weight decay&10$^{-5}$\\
         Video embedding dim&768&dropout&0.5\\
         Image embedding dim&768&hidden dim&256\\
        \bottomrule[1.5pt]
    \end{tabular}
    }
    \caption{Hyperparameter settings of the misinformation detector.}
    \label{tab: hyper}
\end{table}

\subsection{Detector Ablation Study}
\label{subapp: detector ablation study}
We further design various variants of the misinformation detectors, removing certain components to explore which ones are essential for detection. We first remove each component except \emph{content}. Then we design (\romannumeral 1) w/o \emph{Interaction} removing \emph{comment} and \emph{repost}; (\romannumeral 2) w/o \emph{Vison} removing \emph{image} and \emph{video}; and (\romannumeral 3) w/o \emph{extra} only containing \emph{content}. For each variant, we set the remove features as $\mathbf{0}$. For example, if we remove the \emph{comment}, then we set $\boldsymbol{f}_{\textit{comment}}$ as $\mathbf{0}$. We present the ablation study performance in Table \ref{tab: more_main_results}. It illustrates that:
\begin{itemize}[topsep=4pt, leftmargin=*]
    \item The detector without \emph{Extra} modalities suffers a significant performance decline, with an accuracy drop of $17.5\%$. It is more radical, often identifying information as misinformation and achieving high recall. It proves that extra modalities provide valuable signals to identify misinformation. 
    \item The detector without \emph{Interaction} drops to $77.3\%$ on f1-score, illustrating the effectiveness of user reaction including comments and reposts. We speculate that user interactions could provide extra evidence and signals \citep{grover2022public} to verify the information. Meanwhile, reposts provide more evidence than comments. We assume it is related to the algorithm of social platforms, where reposted messages could be spread more widely. Thus users tend to publish verified information when reposting.
    \item The detector w/o \emph{Vision} only drops $2.2\%$ on f1-score, where image and video information could not provide valuable evidence. Meanwhile, video information contributes the least, with the p-value of the t-test on accuracy being $0.015$, which is not considered statistically significant. The text modalities dominate misinformation detection. We speculate that (\romannumeral 1) annotators also consider text information when judging misinformation, introducing biases; and (\romannumeral 2) the pre-trained vision encoders struggle to capture signals related to identifying misinformation.
\end{itemize}

\begin{table}[t]
    \centering
    \resizebox{\linewidth}{!}{
    \begin{tabular}{l|c|c|c|c}
        \toprule[1.5pt]
         Models&Accuracy&F1-score&Precision&Recall\\
         \midrule[1pt]
         Vanilla&$95.2_{\pm 0.6}$&$92.3_{\pm 0.8}$&$93.7_{\pm 1.7}$&$91.0_{\pm 1.0}$\\
         \midrule[1pt]
          ~~w/o \emph{Comment}&$\mathop{93.0^\star_{\pm 1.4}}\limits_{2.3\%\downarrow}$&$\mathop{89.5^\star_{\pm1.6}}\limits_{3.0\%\downarrow}$&$\mathop{86.1^\star_{\pm 3.9}}\limits_{8.1\%\downarrow}$&$\mathop{93.3^\dagger_{\pm 1.6}}\limits_{2.6\%\uparrow}$\\
         ~~w/o \emph{Repost}&$\mathop{89.4^\star_{\pm2.0}}\limits_{6.0\%\downarrow}$&$\mathop{85.3^\star_{\pm 2.2}}\limits_{7.6\%\downarrow}$&$\mathop{76.8^\star_{\pm4.0}}\limits_{18.0\%\downarrow}$&$\mathop{96.1^\star_{\pm 1.4}}\limits_{5.6\%\uparrow}$\\
         ~~w/o \emph{Image}&$\mathop{94.3^\star_{\pm 0.5}}\limits_{1.0\%\downarrow}$&$\mathop{90.5^\star_{\pm 1.0}}\limits_{1.9\%\downarrow}$&$\mathop{95.1^\dagger_{\pm 1.2}}\limits_{1.4\%\uparrow}$&$\mathop{86.5^\star_{\pm 2.0}}\limits_{4.9\%\downarrow}$\\
         ~~w/o \emph{Video}&$\mathop{95.0^\dagger_{\pm0.7}}\limits_{0.2\%\downarrow}$&$\mathop{92.1^\dagger_{\pm 0.8}}\limits_{0.3\%\downarrow}$&$\mathop{93.0^\dagger_{\pm1.9}}\limits_{0.8\%\downarrow}$&$\mathop{91.1^\dagger_{\pm 1.0}}\limits_{0.2\%\uparrow}$\\
         \midrule[1pt]
         ~~w/o \emph{Interaction}&$\mathop{81.6^\star_{\pm4.5}}\limits_{14.2\%\downarrow}$&$\mathop{77.3^\star_{\pm4.1}}\limits_{16.2\%\downarrow}$&$\mathop{64.4^\star_{\pm5.9}}\limits_{31.3\%\downarrow}$&$\mathop{97.3^\star_{\pm 1.2}}\limits_{7.0\%\uparrow}$\\
         ~~w/o \emph{Vision}&$\mathop{94.1^\star_{\pm 0.5}}\limits_{1.1\%\downarrow}$&$\mathop{90.3^\star_{\pm1.0}}\limits_{2.2\%\downarrow}$&$\mathop{94.2^\dagger_{\pm1.2}}\limits_{0.4\%\uparrow}$&$\mathop{86.8^\star_{\pm2.1}}\limits_{4.6\%\downarrow}$\\
         ~~w/o \emph{Extra}&$\mathop{78.5^\star_{\pm5.2}}\limits_{17.5\%\downarrow}$&$\mathop{74.5^\star_{\pm 4.3}}\limits_{19.3\%\downarrow}$&$\mathop{60.6^\star_{\pm6.0}}\limits_{35.4\%\downarrow}$&$\mathop{97.3^\star_{\pm 1.0}}\limits_{6.9\%\uparrow}$\\
         \bottomrule[1.5pt]
    \end{tabular}
    }
    \caption{Performance of the misinformation detector and variants. We report the mean and standard deviation of ten-fold cross-validation. We also report the performance changes and conduct the paired t-test with vanilla, where $\star$ denotes the p-value is less than $0.0005$ and $\dagger$ denotes otherwise. This simple misinformation detector achieves ideal performance. Misinformation and real information are distinguishable with the help of user interactions.}
    \label{tab: more_main_results}
\end{table}

\subsection{Expert Performance}
\label{subapp: expert performance}
We employ 11 social bot detectors as experts. Table \ref{tab: annotator} presents the performance and temperature of these experts. The performance of the automatic annotator is acceptable, proving the credibility of the annotations. Meanwhile, filtering in experts with an accuracy greater than 80\% could improve the annotation precision. To obtain a higher precision, we set the likelihood threshold as 0.75, making sure that the annotator does not identify a genuine account as a social bot (with a precision of 97.6\%).
\begin{table}[]
    \centering
    \resizebox{\linewidth}{!}{
    \begin{tabular}{l|c|c|c|c|c}
        \toprule[1.5pt]
        Experts&Accuracy&F1-score&Precision&Recall&Temperature\\
        \midrule[1pt]

        \midrule[1pt]
        \multicolumn{6}{l}{Feature-based Detectors}\\
        MLP&$73.5$&$49.6$&$90.9$&$34.1$&1.125\\
        Random Forest&$71.7$&$58.0$&$67.0$&$51.1$&$-$\\
        Adaboost&$69.5$&$59.8$&$60.2$&$59.5$&$-$\\
        \midrule[1pt]

        \midrule[1pt]
        \multicolumn{6}{l}{Content-based Detectors (BERT)}\\
        Name&$74.5$&$54.3$&$86.4$&$39.6$&1.468\\
        Description&$75.2$&$56.2$&$86.6$&$41.6$&1.246\\
        Posts$^\star$&$80.4$&$72.4$&$78.4$&$67.2$&1.286\\
        \multicolumn{6}{l}{Content-based Detectors (DeBERTa)}\\
        Name&$74.8$&$54.7$&$87.1$&$39.8$&1.408\\
        Description&$75.2$&$58.7$&$80.9$&$46.1$&0.972\\
        Posts$^\star$&$80.6$&$73.6$&$76.9$&$70.5$&1.129\\
        \midrule[1pt]

        \midrule[1pt]
        \multicolumn{6}{l}{Ensemble Detectors}\\
        BERT$^\star$&$83.1$&$77.3$&$79.4$&$75.4$&1.329\\
        DeBERTa$^\star$&$82.7$&$76.5$&$79.4$&$73.8$&1.146\\
        \midrule[1pt]

        \midrule[1pt]
        Annotator&$85.0$&$79.5$&$83.3$&$76.1$&$-$\\
        All Expert&$82.5$&$72.7$&$89.5$&$61.3$&$-$\\
        Annotator (0.75)&$81.5$&$68.6$&$97.6$&$52.8$&$-$\\
        \midrule[1pt]
        
        \bottomrule[1.5pt]
    \end{tabular}}
    \caption{The performance and temperature of the social bot detectors. The $\star$ indicates that we employ this expert in the final automatic annotator, and $-$ indicates that temperature scaling is not suitable for this expert. The ``All Expert'' denotes the ensemble of all experts. The ``Annotator (0.75)'' denotes that we consider an account a social bot if the likelihood is greater than 0.75.}
    \label{tab: annotator}
\end{table}

\section{Details of Further Analysis}
\subsection{Cluster Algorithm}
\label{subapp: cluster algorithm}
We cluster misinformation into different groups, where each group represents a topic or an event, based on the \textbf{judgment}. The main idea is that the judgments about the same event are very similar. Meanwhile, judgments about distinct events are very different. Formally, we assume the misinformation judgment set is $\{T_i\}_{i=1}^N$, where $N$ is the number of misinformation judgments. Given a specific judgment $T_i$, we calculate the cosine similarities of BERT\footnote{Here is the \href{https://huggingface.co/shibing624/text2vec-base-chinese}{model link}.} representations: 
\begin{align*}
    s_{i,j}=\mathrm{cosine}(\mathrm{BERT}(T_i), \mathrm{BERT}(T_j)).
\end{align*}
We sort the scores $\{s_{i,j}\}_{j=1}^{N}$ in descending order to obtain $\{s_{i,\tilde{j}}\}_{j=1}^{N}$. We then find the index $\hat{j}$ that maximize the gradient:
\begin{align*}
    \hat{j} = \arg\max_{\tilde{j}} s_{i,\tilde{j}}-s_{i,\tilde{j}+1}.
\end{align*}
It means judgments with a similarity score greater than $s_{i,\hat{j}}$ are very similar to $T_i$ and others are very distinct. Here we construct a relation from $T_i$ to the judgments with a similarity score greater than $s_{i,\hat{j}}$. After that, we could obtain a directed graph. We consider each strongly connected graph as a misinformation graph.

\subsection{Top-ten Topics}
\label{subapp: top-ten topics}
Table \ref{tab: topics} presents the keywords and descriptions of the top 10 topics with the highest number of misinformation items. We employ BERT\footnote{Here is the \href{https://huggingface.co/shibing624/text2vec-base-chinese}{model link}.} to obtain the representations of misinformation. 

\begin{table}[]
    \centering
    \resizebox{\linewidth}{!}{
    \small
    \begin{tabular}{l|p{0.7\linewidth}}
        \toprule[1.5pt]
         \textbf{Keyword}&\textbf{Description}\\
         \midrule[1pt]
         Fire Disaster & A place is on fire.\\\midrule[0.5pt]
         Dog Lost Notice& Someone offers a reward of 10 million yuan to find the dog.\\\midrule[0.5pt]
         Import& A country announced a ban on the import of another country's coal.\\\midrule[0.5pt]
         Typhoon& Does it feel like a disaster movie? A place is experiencing a typhoon.\\\midrule[0.5pt]
         Air Crash& The last two minutes of a place's air crash.\\\midrule[0.5pt]
         University Admission& A 19-year-old freshman girl in a city fell to her death and her roommate was recommended for undergraduate study.\\\midrule[0.5pt]
         BBQ& the woman beaten in the barbecue restaurant is dead.\\\midrule[0.5pt]
         Suicide& The woman who jumped from a place had her home disinfected and looted.\\\midrule[0.5pt]
         Child Trafficking& A 5-year-old son in a place was abducted near a bilingual kindergarten.\\\midrule[0.5pt]
         Domestic Violence& The man from a province is the stepfather, and I hope the relevant departments will save this poor child.\\
        \bottomrule[1.5pt]
    \end{tabular}}
    \caption{The keywords and descriptions of 10 topics. We translate them into English and conceal the private information.}
    \label{tab: topics}
\end{table}

\begin{figure*}
    \centering
    \includegraphics[width=\linewidth]{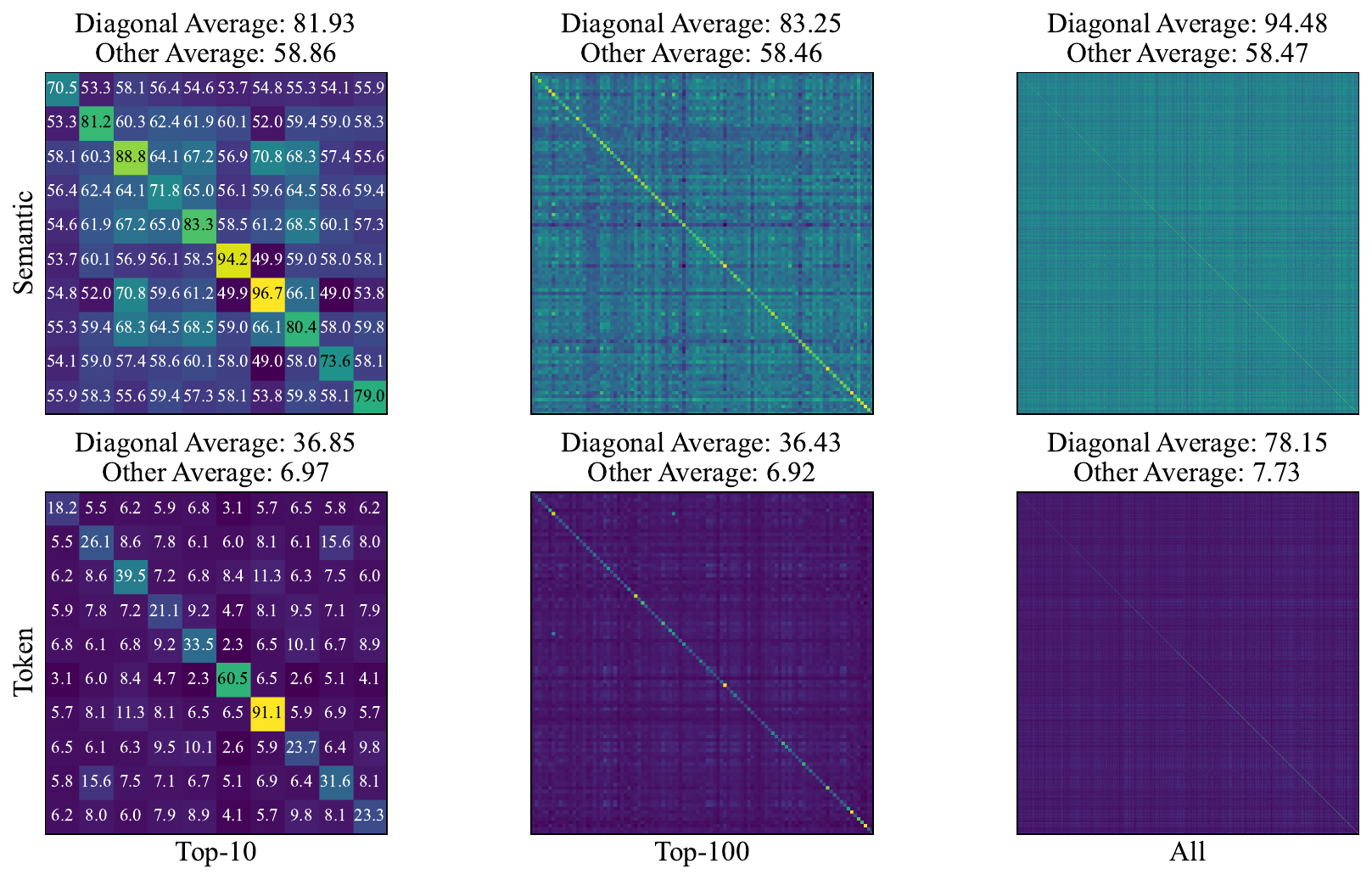}
    \caption{The pairwise score heatmap of \emph{sematic} and \emph{token} levels. The values on the diagonal are significantly larger than the rest. The ``Top-10'' means the 10 topics with the most misinformation instances, the ``Top-100'' means the 100 topics with the most misinformation instances, and the ``All'' means all misinformation instances.}
    \label{fig: heatmap}
\end{figure*}

\subsection{Pairwise Scores}
\label{subapp: pairwise scores}
We conduct numerical analysis to prove that misinformation in the same cluster is similar, while misinformation in different clusters is distinct. We employ \emph{semantic-level} and \emph{token-level} pairwise scores. Formally, we assume there are $N$ clusters (2,270 clusters), and the $i$-th cluster is represented as $\{T_k^i\}_{i=k}^{M_i}$, where $M_i$ if the number of misinformation in this cluster. Given the $i$-th cluster and $j$-th cluster, the pairwise score $s_{ij}$ is calculated as:
\begin{align*}
    s_{ij} = \frac{1}{M_iM_j}\sum_{p=1}^{M_i}\sum_{q=1}^{M_j}\mathrm{score}(T_p^i,T_q^j),
\end{align*}
where $\mathrm{score}(\cdot, \cdot)$ is the similarity function. For \emph{semantic}, we employ the cosine similarity of BERT\footnote{Here is the \href{https://huggingface.co/shibing624/text2vec-base-chinese}{model link}.} representation. For \emph{token}, we employ the jieba package\footnote{\url{https://pypi.org/project/jieba/}} to tokenize Chinese sentences and calculate the ROUGE-L score. Since computing pairwise ROUGE-L is time-consuming, we randomly sample 10 pieces of misinformation in each cluster.

\begin{figure}[t]
    \centering
    \includegraphics[width=\linewidth]{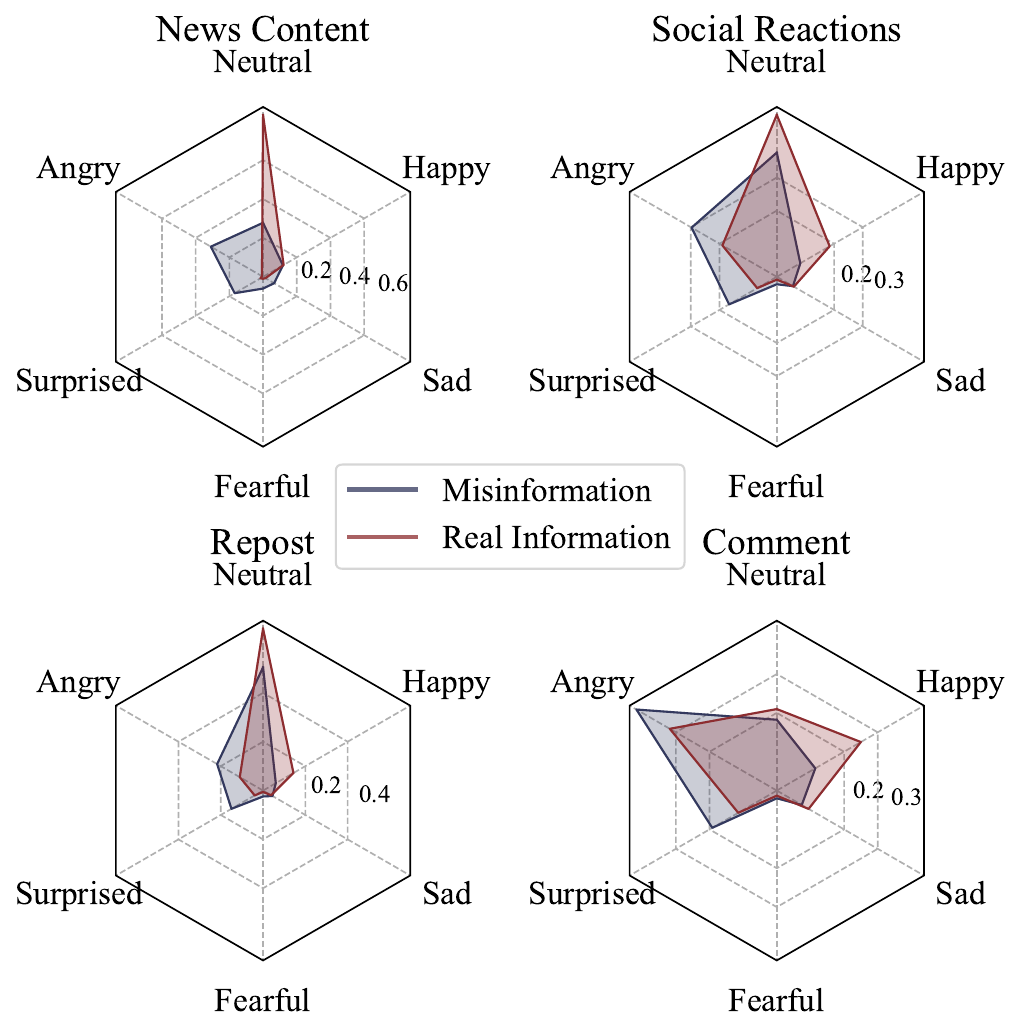}
    \caption{Sentiment distributions in different texts. Misinformation would publish emotional content while real information would publish more neutral content. Users would publish emotional content during the information spread.}
    \label{fig: sentiment}
\end{figure}

\subsection{Score heatmap}
\label{subapp: score heatmap}
Figure \ref{fig: heatmap} presents the heatmap of the pairwise score, which illustrates that the values in the diagonal are much greater. It enhances our findings that: misinformation with the same topics has similar content and misinformation with different topics has distinct content. 
\subsection{Sentiments and Stances}
\label{subapp: sentiments and stances}
To obtain the sentiments of social texts, we employ BERT trained on the EWECT dataset\footnote{\url{https://smp2020ewect.github.io/}}. The sentiments include \emph{neutral}, \emph{happy}, \emph{angry}, \emph{surprised}, \emph{sad}, and \emph{fearful}. To obtain the stances of social texts, we employ BERT trained on the STANCE dataset \citep{zhao2023c}. The stances include \emph{support}, \emph{oppose}, and \emph{neutral}.

\subsection{Sentiment Distribution}
\label{subapp: sentiment distribution}
Figure \ref{fig: sentiment} illustrates the distribution of sentiments in different texts. An intuitive finding is that misinformation would publish more emotional content while real news would naturally report. However, whether in misinformation or real news, public reactions are always emotional. Comments in misinformation show more anger while real news shows more happiness, both of which are more emotional than reposts. We speculate that users are inclined to comment to express emotion.

\subsection{Sentiment Variation}
\label{subapp: sentiment variation}
We introduce the variation measure to calculate the degree or extent to which public sentiment changes over the news spread. Given a specific information instance and its relation comment, we first calculate the function of the proportion of comments with neutral sentiment over time $f(x)$. We then determine the time series $[x_0, x_1, \dots, x_n]$, where we set the interval as one hour. The variation is calculated as:
\begin{align*}
    v_{\Delta}=\sum_{k=1}^n|f(x_k)-f(x_{k-1})|.
\end{align*}
\subsection{Correlation Coefficient}
\label{subapp: correlation coefficient}
To numerically explore the correlations between social bots and online public opinions, we calculate the following Pearson correlation coefficient:
\begin{itemize}[topsep=4pt, leftmargin=*]
    \item The number of social bots and the number of comments with non-neural stances: 0.6661.
    \item The number of social bots and the number of comments with non-neural sentiments: 0.6750.
    \item The ratio of social bots and the ratio of comments with non-neural stances: 0.2040.
    \item The ratio of social bots and the ratio of comments with non-neural sentiments: 0.2499.
\end{itemize}

The relatively high correlation coefficients indicate that social bots might influence public opinion.
\subsection{Semantic Similarity}
\label{subapp: semantic similarity}
We explore the publishing behavior differences between social bots and genuine accounts. Here we explore whether accounts would publish similar content by introducing the semantic similarity score. Given an account with its posts in the timeline $\{T_i\}_{i=1}^N$, the semantic similarity is calculated as:
\begin{align*}
    s = \frac{1}{N^2}\sum_{i=1}^N\sum_{j=1}^N\mathrm{cosine}(\mathrm{BERT}(T_i), \mathrm{BERT}(T_j)).
\end{align*}



\end{document}